\documentclass[12pt]{article}

\usepackage{algpseudocode}
\usepackage{algorithm}
\usepackage{amsmath,amssymb}
\usepackage{url,epsfig}
\usepackage{fancyvrb}
\usepackage{geometry}
\usepackage{sectsty}
\usepackage[normalem]{ulem}
\usepackage{breqn}
\usepackage{hyperref}
\sectionfont{\sffamily\bfseries\upshape\large}
\subsectionfont{\sffamily\bfseries\upshape\normalsize}
\subsubsectionfont{\sffamily\mdseries\upshape\normalsize}
\makeatletter
\renewcommand\@seccntformat[1]{\csname the#1\endcsname.\quad}

\makeatletter
\def\@maketitle{%
  \begin{center}%
  \let \footnote \thanks
    {\large \@title \par}%
    {\normalsize
      \begin{tabular}[t]{c}%
        \@author
      \end{tabular}\par}%
    {\small \@date}%
  \end{center}%
}
\makeatother

\makeatletter
\providecommand*{\diff}%
{\@ifnextchar^{\DIfF}{\DIfF^{}}}
\def\DIfF^#1{%
\mathop{\mathrm{\mathstrut d}}%
\nolimits^{#1}\gobblespace}
\def\gobblespace{%
\futurelet\diffarg\opspace}
\def\opspace{%
\let\DiffSpace\!%
\ifx\diffarg(%
\let\DiffSpace\relax
\else
\ifx\diffarg[%
\let\DiffSpace\relax
\else
\ifx\diffarg\{%
\let\DiffSpace\relax
\fi\fi\fi\DiffSpace}

\title{\bf Voting patterns in 2016: Exploration using multilevel regression and
poststratification (MRP) on pre-election polls \vspace{.5\baselineskip}}

\author{Rob Trangucci\thanks{University of Michigan} \footnotemark[3]
\and Imad Ali\footnotemark[2]
\and Andrew Gelman\thanks{Columbia University}
\and Doug Rivers\thanks{YouGov} \thanks{Stanford University}}

\date{\vspace{.5\baselineskip} 01 February 2018}

\begin{document}

\maketitle

\begin{abstract}\noindent
We analyzed 2012 and 2016 YouGov pre-election polls in order to understand
how different population groups voted in the 2012 and 2016 elections. We
broke the data down by demographics and state and found:
\begin{itemize}
\item The gender gap was an increasing function of age in 2016.
\item In 2016 most states exhibited a U-shaped gender gap curve with respect to
  education indicating a larger gender gap at lower and higher levels of
    education.
\item Older white voters with less education more strongly supported Donald
  Trump versus younger white voters with more education.
\item Women more strongly supported Hillary Clinton than men, with young and
  more educated women most strongly supporting Hillary Clinton.
\item Older men with less education more strongly supported Donald Trump.
\item Black voters overwhelmingly supported Hillary Clinton.
\item The gap between college-educated voters and non-college-educated voters
  was about 10 percentage points in favor of Hillary Clinton
\end{itemize}
We display our findings with a series of graphs and maps. The R code
  associated with this project is available at
  \url{https://github.com/rtrangucci/mrp_2016_election/}.
\end{abstract}

\newpage
\tableofcontents
\newpage

\section{Introduction}

After any election, we typically want to understand how the electorate voted.
While national and state results give exact measures of aggregate voting, we
may be interested in voting behavior that cuts across state lines, such as how
different demographic groups voted. Exit polls provide one such measure, but
without access to the raw data we cannot determine aggregates beyond the
margins that are supplied by the exit poll aggregates.

In pursuit of this goal, we can use national pre-election polls in which
respondents are asked for whom they plan to vote and post-election polls in
which respondents are asked if they participated in the election, both of which
record demographic information and state residency of respondents. Using this
data, we then build a statistical model that uses demographics and state
information to predict the probability that an eligible voter voted in the
election and which candidate a voter supports. A model that accurately predicts
voting intentions for specific demographic groups (e.g.  college-educated
Hispanic men living in Georgia) will require deep interactions as outlined in
\cite{ghitza2013deep}. In order to precisely learn the second- and third-order
interactions, we require a large dataset that covers many disparate groups.

Armed with our two models, we can use U.S. Census data to yield the number of
people in each demographic group. For each group, we then predict the number of
voters, and the number of votes for each candidate to yield a fine-grained
dataset. We can then aggregate this dataset along any demographic axes we
choose in order to investigate voting behavior. 

\section{Data and methods}

\subsection{Data}

We use YouGov's daily tracking polls from 10/24/2016 through 11/6/2016 to train
the 2016 voter preference model. We included 56,946 respondents in the final
dataset after filtering out incomplete cases. To train the 2012 voter
preference model we used 18,716 respondents polled on 11/4/2012 from YouGov's
daily tracking poll.

In order to train the 2016 voter turnout model, we use the Current Population
Survey (CPS) from 2016, which includes a voting supplement (\cite{ipums}). The model used
80,766 responses from voters as to whether they voted in the 2016 presidential
election. We used the CPS from 2012 to train the 2012 voter turnout model,
which comprises 81,017 voters. We decided to use the CPS to train our model
because it is viewed as the gold-standard in voter-turnout polling
\cite{lei20162008}.

We use a modified version of the 2012 Public Use Microdata Sample Census
dataset (PUMS) to get a measure of the total number of eligible voters in the
U.S. YouGov provided the PUMS dataset with ages and education adjusted to match the
2016 population.

\subsection{Methods}

Our methodology follows that outlined in \cite{gelman1997poststratification},
\cite{ghitza2013deep}, and \cite{lei20162008}. For voter $i$ in group $g$ as
defined by the values of a collection of categorical variables, we want
to learn the voter's propensity to vote and for whom they plan to vote, by
using a nonrandom sample from the population of interest. We assume that an
individual voter's response in group $g$ is modeled as follows:
\begin{align*}
  T_i \sim \text{Bernoulli}(\alpha_{g[i]})
\end{align*}
where $T_i$ is $1$ if the voter plans to vote for Trump, or $0$ otherwise.
$\alpha_{g[i]}$ is the probability of voting for Trump for voter $i$ in group
$g$. In order to make inferences about $\alpha_{g[i]}$ without modeling the
selection process, we need to stratify our respondents into small enough groups
so that within a cell selection is random (i.e. that the responses are
Bernoulli random variables conditional only on $g$). We do so by generating
multidimensional cells defined by demographic variables like age, ethnicity,
and state of residence that categorize our respondents. This induces data
sparsity even in large polls so we must use Bayesian hierarchical models to
partially pool cells along these demographic axes.

Upon fitting our model, we can use the posterior mean of $\alpha_g$,
$\hat{\alpha}_g$ and Census data to estimate an aggregate Trump vote proportion
by calculating the weighted average $\sum_{g \in D}\tfrac{N_g
\hat{\alpha}_g}{N_D}$ for whatever demographic category $D$ we like.

We measure our electorate using six categorical variables:
\begin{itemize}
  \item State residency
  \item Ethnicity
  \item Gender
  \item Marital status
  \item Age
  \item Education
\end{itemize}
Each variable $v$ has $L_v$ levels. State residency has fifty levels.
Ethnicity has four levels: Black, Hispanic, Other, and White. Gender has two
levels.  Marital status has three levels: Never married, Married, Not
married. Age has four levels, corresponding to the left-closed intervals of
age: $[18,30), [30,45), [45,65), [65,99)$.  Education has five levels: No High
School, High School, Some College, College, Post Graduate.

After binning our Census data by the six-way interaction of the above
attributes, we generate table \ref{ps_table}.  Each row of the table represents
a specific group of the population, an intersection of six observable
attributes. We refer to each row as a cell, and the full table as a six-way
poststratification table. Our table has 33,561 cells, reflecting the fact that
not all possible six-way groups exist in the U.S..

\begin{table}[ht] \label{ps_table}
\centering
\caption{Six-way poststratification table}
\begin{tabular}{rrrrrrrrrr}
  \hline
  Cell index $g$ & State & Ethn. & Gender & \dots & Educ. & N & $\phi_g$ & $\alpha_g | \text{vote}$ & $\mathbb{E}\left[\text{T}_g\right]$ \\
  \hline
  1 & AK  & Black & Female & \dots & College & 400 & 0.40 & 0.50 & 80\\
  2 & AK & Black & Female & \dots & High School & 300 & 0.30 & 0.60 & 54 \\
	\dots & \dots	 & \dots & \dots & \dots & \dots & \dots & \dots \\
	\dots	 & \dots & \dots & \dots & \dots & \dots & \dots & \dots\\
	33651 & WY  & White & Male & \dots & Some College & 200 & 0.40 & 0.40 & 32 \\
   \hline
\end{tabular}
\end{table}

We then add columns to this dataset that represent the cell-by-cell probability
of voting and the cell-by-cell probability of supporting Trump, which can be
combined to yield the expected number of Trump voters,
$\mathbb{E}\left[\text{T}_g\right]$, in each cell $g$:
$\mathbb{E}\left[\text{T}_g\right] = N \times \phi_g \times \alpha_g  |
\text{vote} $
where $\phi_g$ is the expected probability of voting in cell $g$,
and $\alpha_g | \text{vote}$ is the expected probability of voting for Trump for voters in cell $g$

In order to generate $\phi_g$ and $\alpha_g | \text{vote}$, we build two
models: a voter turnout model and a vote preference model, respectively. Both
models are hierarchical binomial logistic regression models of the form:

\begin{align*}
	T_g & \sim \text{Binomial}(V_g, \phi_g)\, , \, g \in \{1,\dots, G\}\\
	\text{logit}\,\phi_g & = \mu + \sum_{v \, \in \, V}
	\beta^v_{\left[v[g]\right]} \\
	\beta^v & \sim \text{Normal}(0, \tau_v)\, \forall v \, \in \, V \\
  \tau_v & = \sqrt{\pi_v |V| S^2} \\
  \boldsymbol{\pi} & \sim \text{Dirichlet}(\mathbf{1}) \\
  S & \sim \text{Gamma}(1, 1)
\end{align*}

Each categorical predictor, $\beta^v$, is represented as a length-$L_v$
vector, where the elements of the vector map to the effect associated with the
level $l_v$. $V$ denotes the set of all categorical predictors included in the
model and $v[g]$ is a function that maps the $g$-th cell to the appropriate
$l_v$-th level of the categorical predictor. For example, $\beta^{\text{state}}$
would be a 50-element vector, and $\text{state}[\,\,]$ is a length-$G$ list of
integers with values between 1 and 50 indicating to which state the $g$-th cell 
belongs. Note that the model above can include one-way effects in $V$, as well as
two-way and three-way interactions, like state $\times$ age.

We use \texttt{rstanarm} to specify the voter turnout model and the voter
preference model, which uses \texttt{lme4} syntax to facilitate building
complex hierarchical generalized linear models like above. The full model
specifications in $\texttt{lme4}$ syntax are given in the Appendix.
\texttt{rstanarm} imposes more structure on the variance parameters $\tau_v$
than is typical. In our model, $\tau_v^2$ is the product of the square of a
global scale parameter $S$ the $v$-th entry in the simplex parameter
$\boldsymbol{\pi}$, and the cardinality of $V$, $|V|$.  See \cite{rstanarm}
for more details.

Our voter preference model went through multiple iterations before we arrived
at our final model. At first we intended to include past presidential vote.
However, PUMS does not include past presidential vote, so we used YouGov's
imputed past presidential vote for each PUMS respondent. This induced too much
sparsity in our poststratification frame.

After training each of the models, and generating predictions for voter turnout
by cell and two-party vote preference for each cell, we adjusted our turnout
and vote proportions in each cell to match the actual state-by-state outcomes
as outlined \cite{ghitza2013deep}.

\begin{table}[H]
\caption{Variables in the vote preference model}
\footnotesize
\begin{center}
\begin{tabular}{ l l l c }
 \texttt{stan\_glmer()} Variable & Description & Type &  Number of Groups \\
 \hline
y & Vote choice & Outcome variable & - \\
1 & Intercept & Global intercept & - \\
female & Fem.: 0.5, Male: -0.5 & Global slope & - \\
state\_pres\_vote & Pre-election poll average  & Global slope & - \\
state & State of residence & Varying intercept & 50 \\
age & Age & Varying intercept & 4 \\
educ & Education attained & Varying intercept & 5 \\
1 + state\_pres\_vote $|$ eth & Ethnicity  & Varying intercept and slope & 4 \\
marstat & Marital status & Varying intercept & 3 \\
marstat:age &  & Varying intercept & 3$\times$4 = 12 \\
marstat:state &  & Varying intercept & 3$\times$50 = 150 \\
marstat:eth &  & Varying intercept & 3$\times$4 = 12 \\
marstat:gender &  & Varying intercept & 3$\times$2 = 6 \\
marstat:educ &  & Varying intercept & 3$\times$5 = 15 \\
state:gender &  & Varying intercept & 50$\times$2 = 100 \\
age:gender &  & Varying intercept & 4$\times$2 = 8\\
educ:gender &  & Varying intercept & 5$\times$2 = 10 \\
eth:gender &  & Varying intercept & 4$\times$2 = 8 \\
state:eth &  & Varying intercept & 50$\times$4 = 200 \\
state:age &  & Varying intercept & 50$\times$4 = 200\\
state:educ &  & Varying intercept & 50$\times$5 = 250 \\
eth:age &  & Varying intercept & 4$\times$4 = 16 \\
eth:educ &  & Varying intercept & 4$\times$5 = 20 \\
age:educ &  & Varying intercept & 4$\times$5 = 20 \\
state:educ:age &  & Varying intercept & 50$\times$4$\times$4 = 800\\
educ:age:gender &  & Varying intercept & 5$\times$4$\times$2 = 40\\
\hline \\
\end{tabular}
\end{center}
\end{table}

\section{Results}

This section presents plots at the county and state level, followed by charts
and maps that illustrate the poststratification. In addition to vote intention,
the charts and maps also illustrate voter turnout. The county and state level
plots use 2016 and 2012 election results and 2010 US census data. The captions
of the charts and maps identify which model is used to produce the data
illustrated in the figure. The models are defined as follows:

\begin{description}
\item[Model 1] is described in \emph{Section 2} above.
\item[Model 2] is similar to \emph{Model 1} but includes income as a factor variable and omits marital status. The 2016 vote turnout model for Model 2 was fitted to 2012 CPS.
\end{description}

\subsection{Election results graphs}

The graphs that follow present actual election results by county and by state.
They are not model-based, but rather an examination of the Republican vote
proportion swing from 2012 to 2016 by county versus various demographic
variables measured at the county level.

\subsubsection{County-level vote swings}

\begin{figure}[H]\caption[]{County-level Republican Swing by Income}
\begin{center}
\includegraphics[trim={0cm 0cm 0cm 4cm}, clip, scale=0.3]{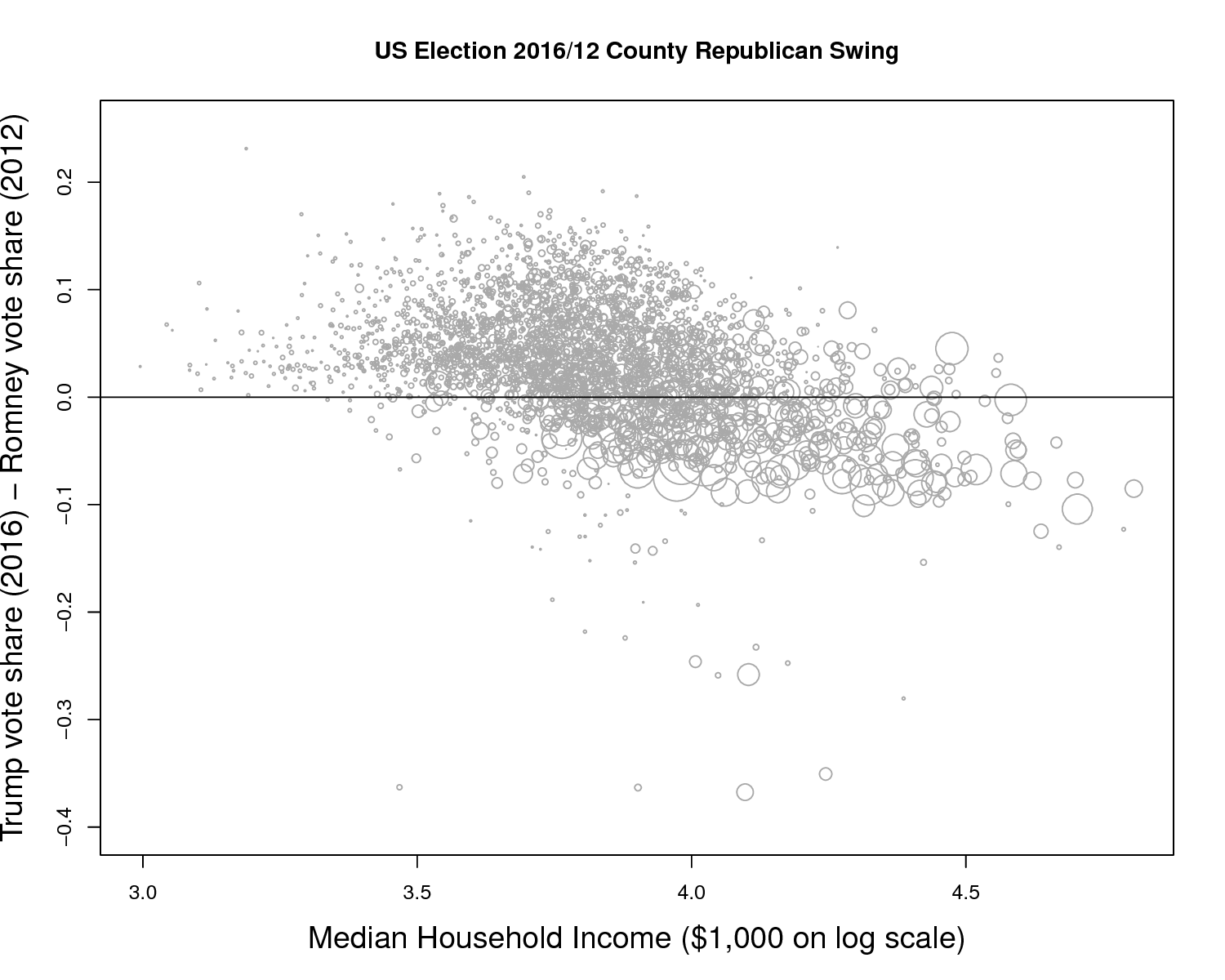}
\end{center}
\footnotesize
\emph{Notes:} The county-level Republican swing is computed as Donald Trump's 2016 two-party vote share minus Mitt Romney's 2012 two-party vote share. Positive values indicate Trump outperforming Romney, while negative values indicate Romney outperforming Trump. The area of each circle is proportional to the number of voters in each county. Overall, Trump outperformed Romney in counties with lower median income. While Trump mostly outperformed Romney in counties with lower voter turnout, Romney mostly outperformed Trump in counties with larger voter turnout.
\end{figure}
\begin{figure}[H]\caption[]{County-level Republican Swing by College Education}
\begin{center}
\includegraphics[trim={0cm 0cm 0cm 4cm}, clip, scale=0.3]{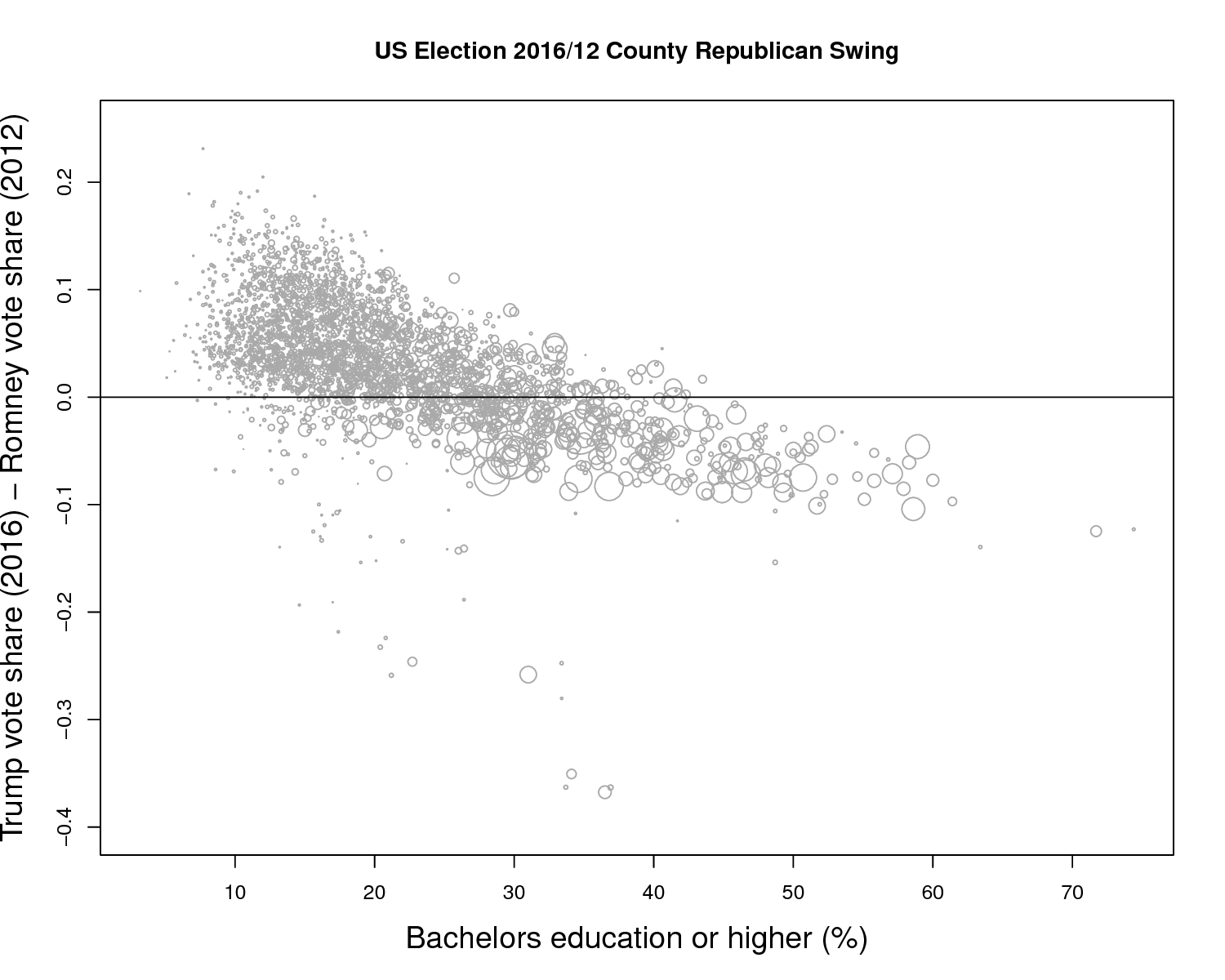}
\end{center}
\footnotesize
\emph{Notes:} The county-level Republican swing is computed as Donald Trump's 2016 two-party vote share minus Mitt Romney's 2012 two-party vote share. Positive values indicate Trump outperforming Romney, while negative values indicate Romney outperforming Trump. The area of each circle is proportional to the number of voters in each county. Overall, Trump outperformed Romney in counties with lower college education. While Trump mostly outperformed Romney in counties with lower voter turnout, Romney mostly outperformed Trump in counties with larger voter turnout.
\end{figure}
\begin{figure}[H]\caption[]{County-level Republican Swing by Region as a Function of Income and College Education}
\begin{center}
\includegraphics[trim={0cm 0cm 0cm 4cm}, clip, scale=0.18]{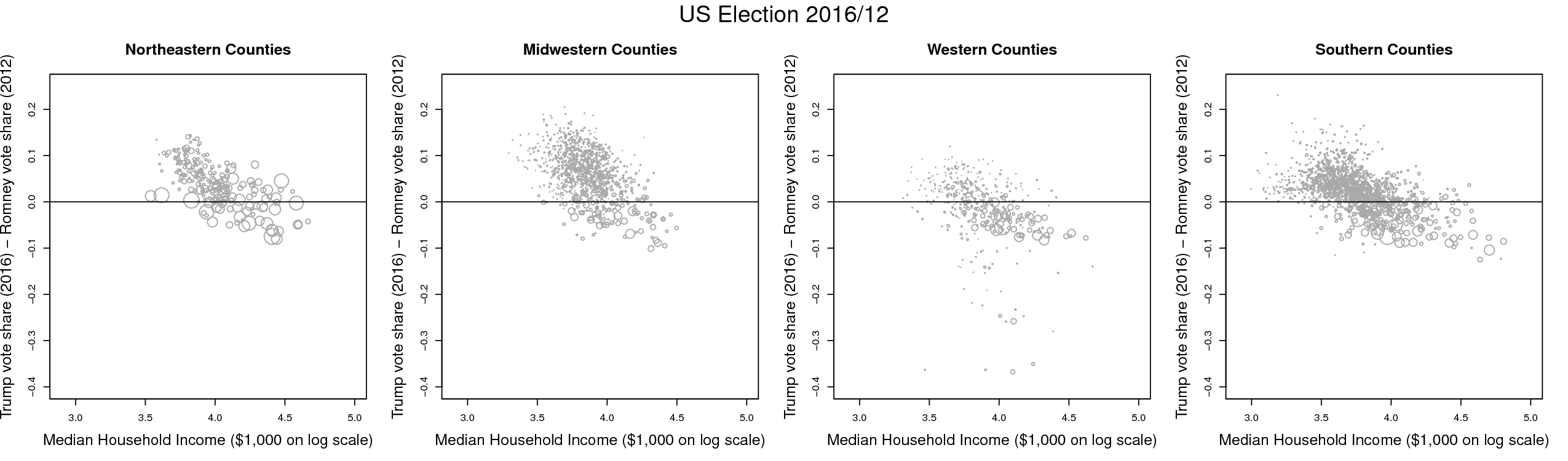}
\includegraphics[trim={0cm 0cm 0cm 4cm}, clip, scale=0.2]{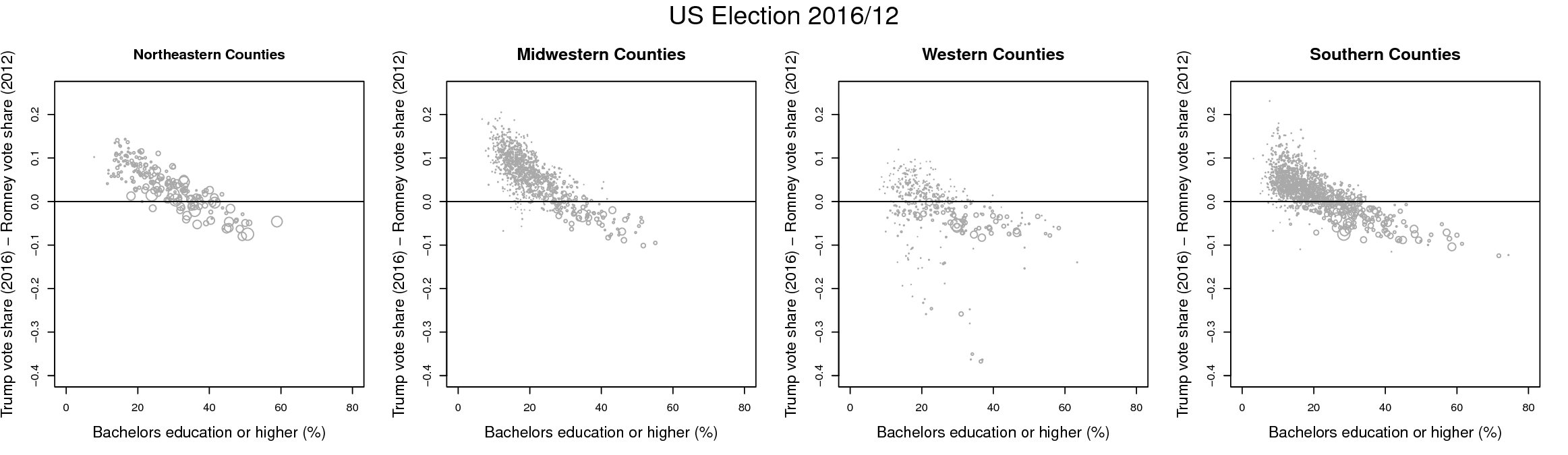}
\end{center}
\footnotesize
\emph{Notes:} The county-level Republican swing is computed as Donald Trump's 2016 two-party vote share minus Mitt Romney's 2012 two-party vote share. Positive values indicate Trump outperforming Romney, while negative values indicate Romney outperforming Trump. The area of each circle is proportional to the number of voters in each county. Across all regions there is a trend of Trump outperforming Romney in low income counties and counties with lower college education. The trend of Trump performing well in counties with lower college education is less apparent in western counties.
\end{figure}

\subsubsection{State-level election results and vote swings}

\begin{figure}[H]\caption[]{Republican Share of the Two-Party Vote 2012-2016}
\begin{center}
\begin{minipage}{0.6\linewidth}
\includegraphics[trim={0cm 0cm 0cm 1cm}, clip, scale=0.6]{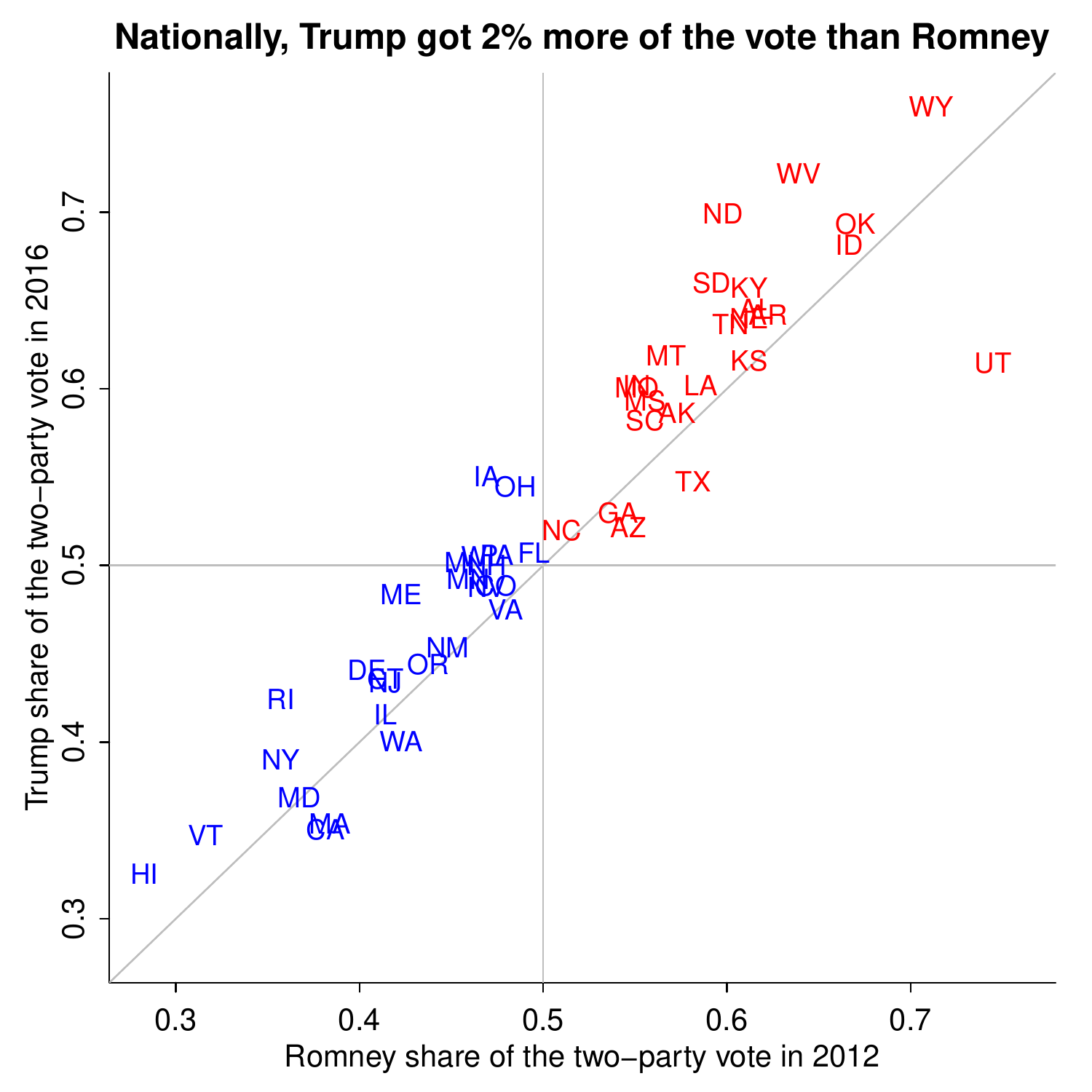}
\footnotesize
\emph{Notes:} The state-level Republican share of the two-party vote. States are color coded according to the results of the 2012 election. States won by Mitt Romney are in red and states won by Barack Obama are in blue. The diagonal line indicates that the 2012 and 2016 Republican candidates received identical shares of the two-party vote. In most states Trump received a higher share of the two-party vote. Nationally, Trump got 2 percent more of the two-party vote than Romney.
\end{minipage}
\end{center}
\end{figure}
\begin{figure}[H]\caption[]{Republican Swing from 2012 to 2016}
\begin{center}
\begin{minipage}{0.75\linewidth}
\includegraphics[trim={0cm 0cm 0cm 1cm}, clip, scale=0.6]{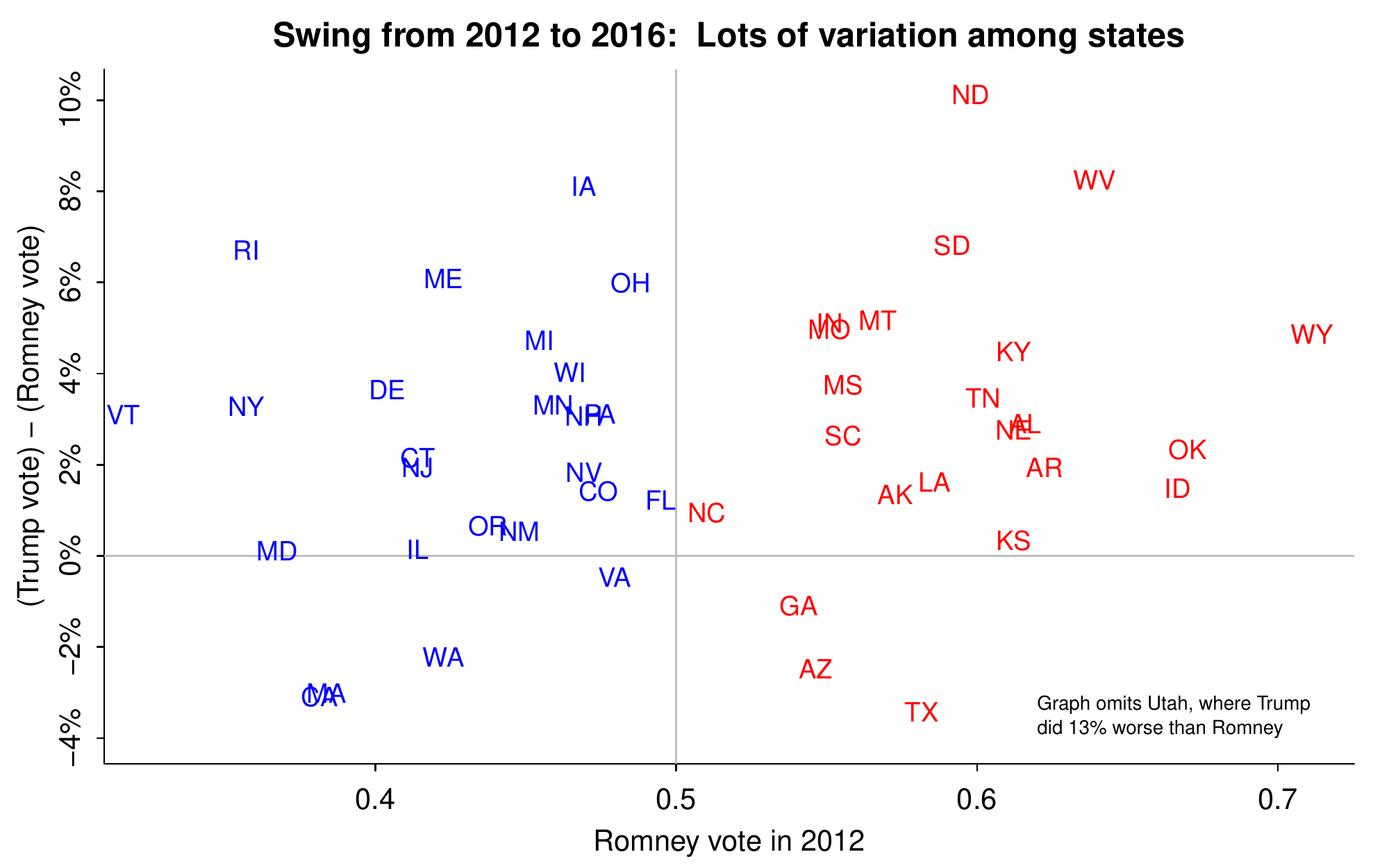}
\footnotesize
\emph{Notes:} The state-level Republican swing. States are color coded according to the results of the 2012 election. States won by Mitt Romney are in red and states won by Barack Obama are in blue. Positive values indicate Trump outperforming Romney and negative values indicate Romney outperforming Trump. There is lots of variation among states with Trump outperforming Romney in most states.
\end{minipage}
\end{center}
\end{figure}
\begin{figure}[H]\caption[]{Trump's Actual and Forecasted Vote Share}
\begin{center}
\begin{minipage}{0.6\linewidth}
\includegraphics[trim={0cm 0cm 0cm 1cm}, clip, scale=0.6]{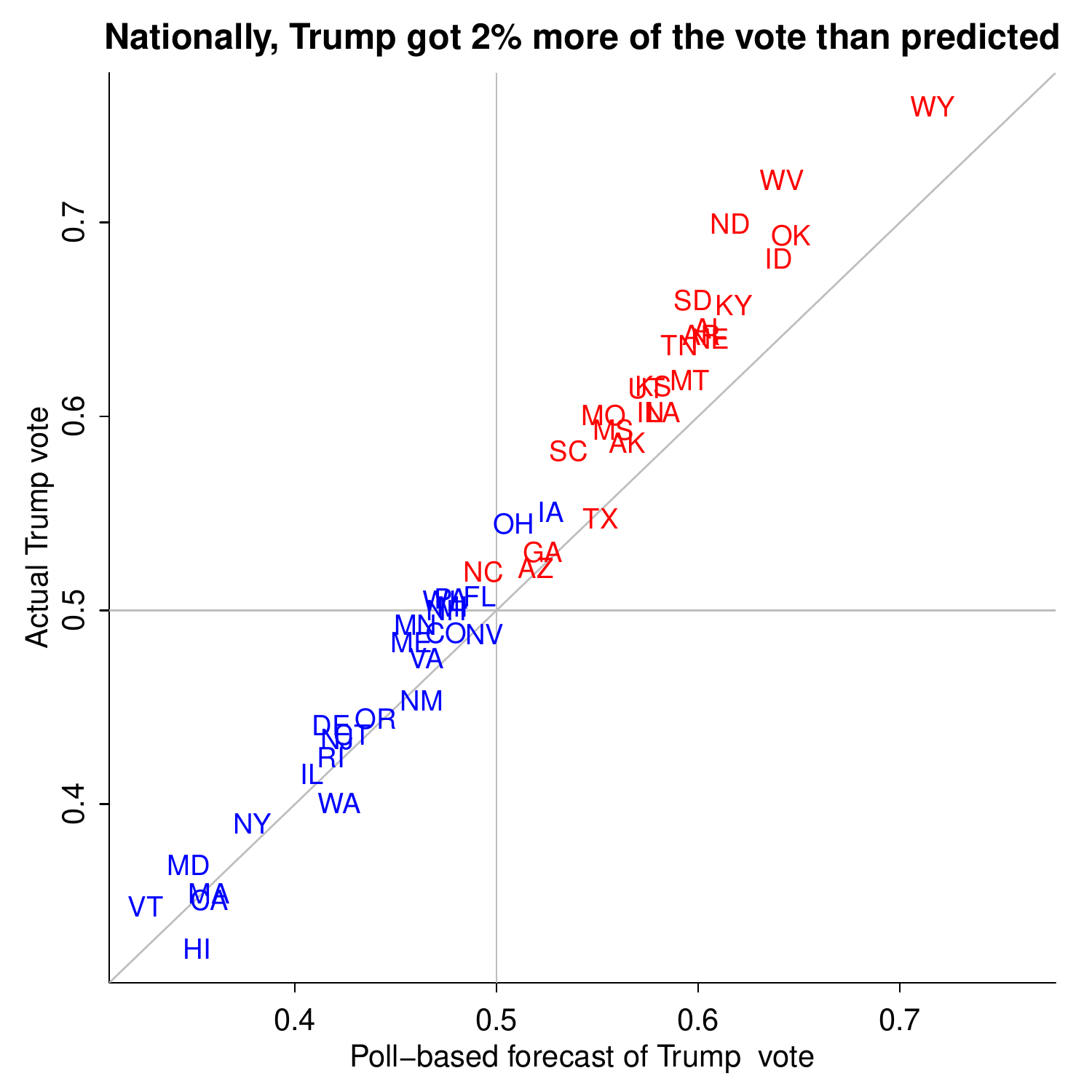}
\footnotesize
\emph{Notes:} A state-level comparison between Donald Trump's actual two-party vote share and his forecasted vote share. States are color coded according to the results of the 2012 election. States won by Mitt Romney are in red and states won by Barack Obama are in blue. Values on the diagonal indicate that Trump's actual performance was in line with his forecast. In most states Trump outperformed his poll-based forecast.
\end{minipage}
\end{center}
\end{figure}
\begin{figure}[H]\caption[]{Trump's Actual Minus Forecasted Vote Share}
\begin{center}
\begin{minipage}{0.75\linewidth}
\includegraphics[trim={0cm 0cm 0cm 1cm}, clip, scale=0.6]{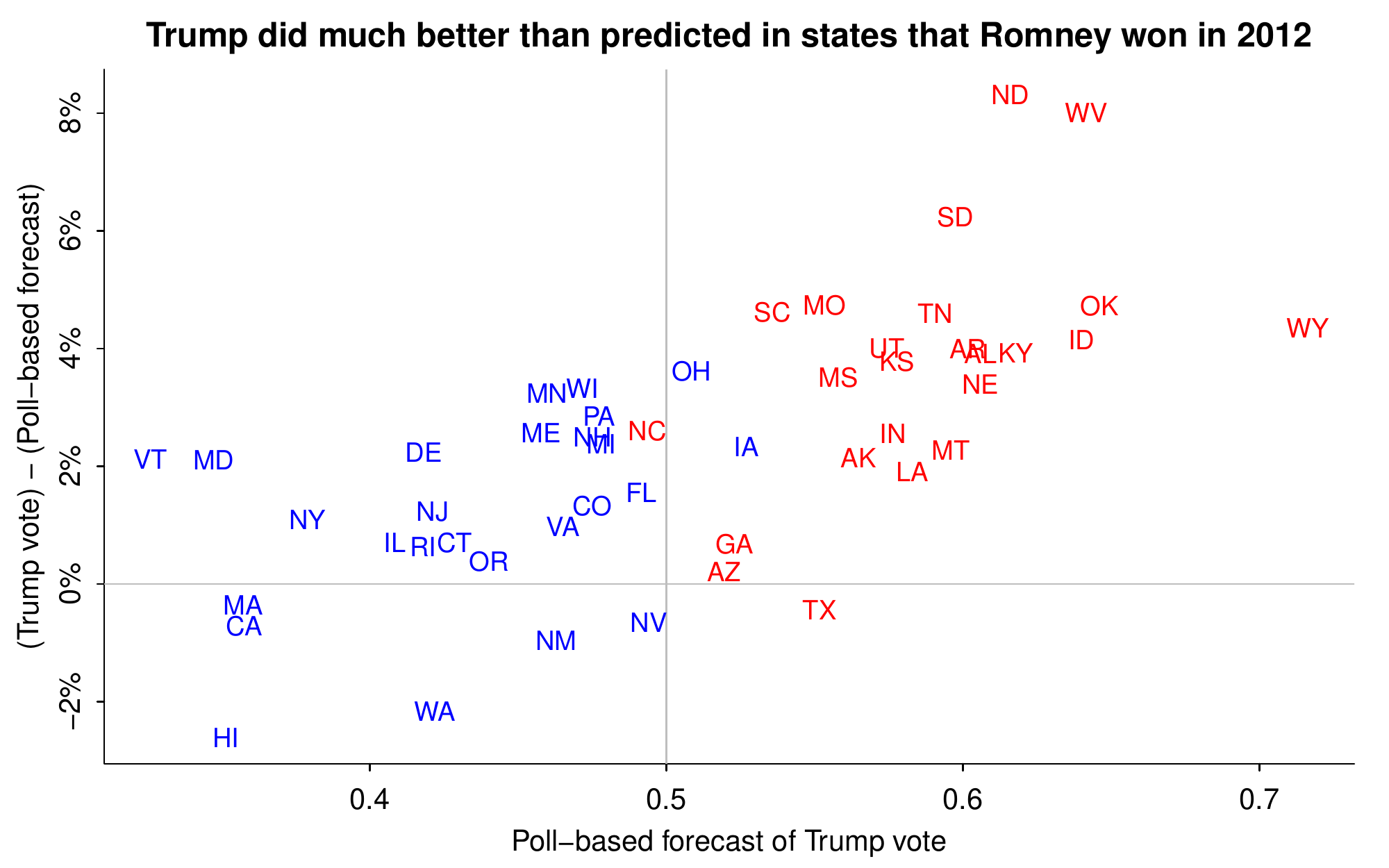}
\footnotesize
\emph{Notes:} A state-level comparison of Donald Trump's actual vote share against his poll-based forecast. States are color coded according to the results of the 2012 election. States won by Mitt Romney are in red and states won by Barack Obama are in blue. Positive values indicate states in which Trump outperformed his forecast and negative values indicate in which Trump's actual performance fell behind his forecast.  Trump did better than predicted in states that Romney won in 2012.
\end{minipage}
\end{center}
\end{figure}

\subsection{Poststratification graphs}

The graphs that follow are generated using the multilevel regression and poststratification
method outlined in the Methodology section.

\subsubsection{Gender gap}
\begin{figure}[H]\caption[]{Gender Gap (Men minus Women) by Education and Age}
\begin{center}
\begin{minipage}{0.75\linewidth}
\includegraphics[trim={0cm 3cm 0cm 1.5cm}, clip, scale=0.4]{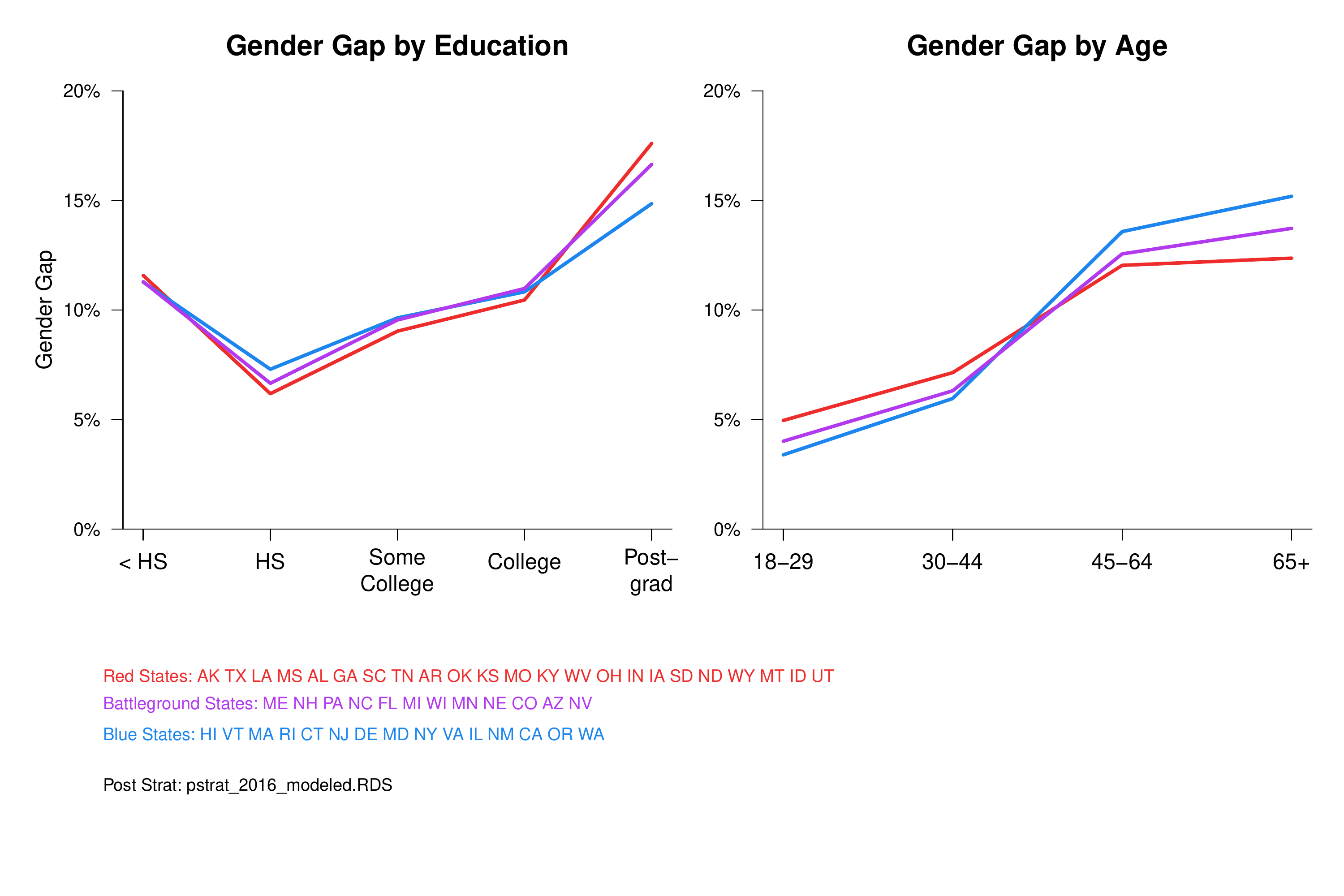}
\footnotesize
\emph{Notes:} The gender gap is evaluated as men's probability of voting for Trump minus women's probability for of voting for Trump for various education and age levels. Larger values indicate a greater divergence in vote preference between men and women. \\
(Using \emph{Model 1}.)
\end{minipage}
\end{center}
\end{figure}
\begin{figure}[H]\caption[]{Gender Gap (Men minus Women) by Education and Age - 2012 Election}
\begin{center}
\begin{minipage}{0.75\linewidth}
\includegraphics[trim={0cm 3cm 0cm 1.5cm}, clip, scale=0.4]{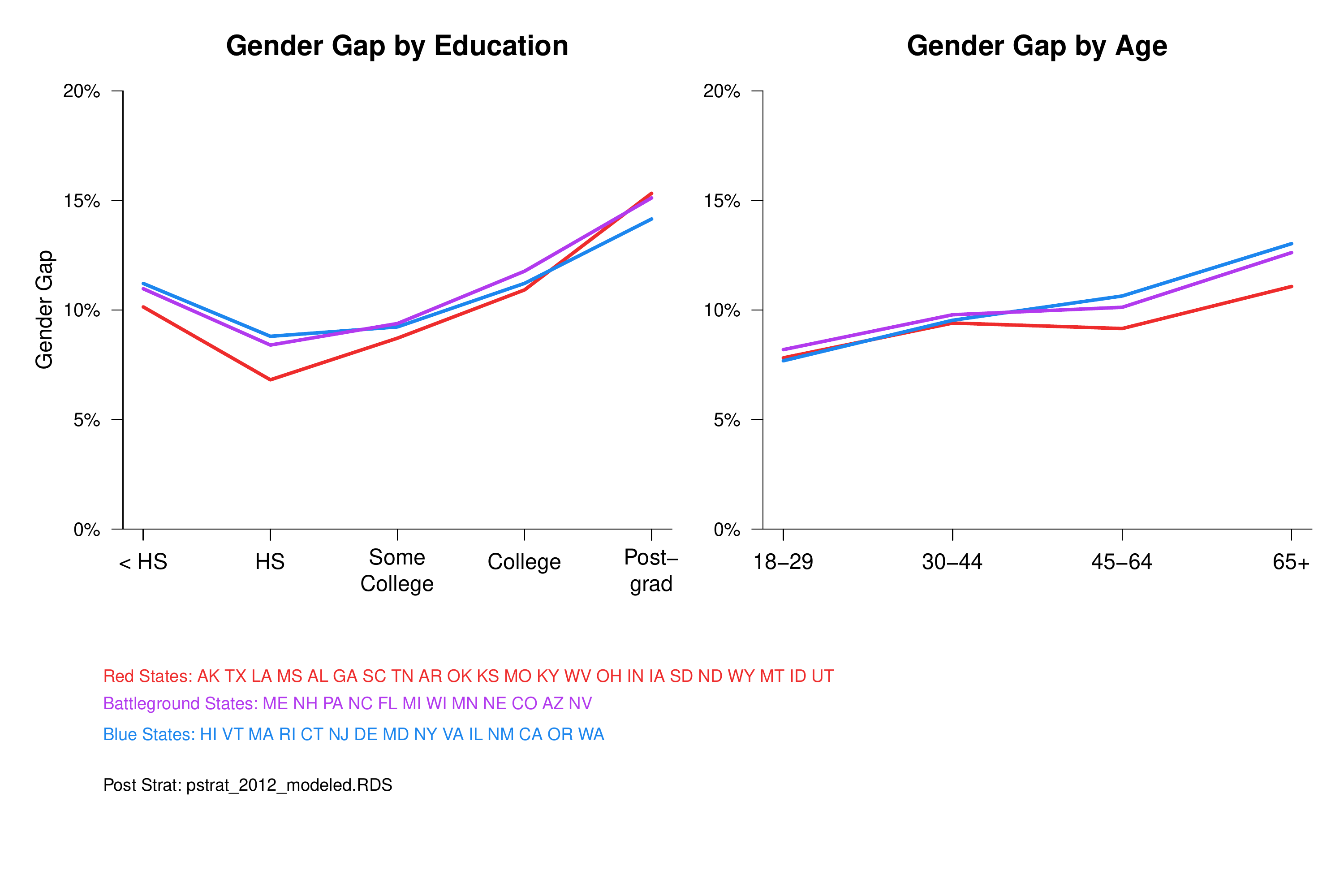}
\footnotesize
\emph{Notes:} The gender gap is evaluated as men's probability of voting for Romney minus women's probability for of voting for Romney for various education and age levels. \\
(Using \emph{Model 1} with 2012 election results/turnout data.)
\end{minipage}
\end{center}
\end{figure}
\begin{figure}[H]\caption[]{Gender Gap by Education for each Age Category}
\begin{center}
\begin{minipage}{0.95\linewidth}
\includegraphics[trim={0cm 2.5cm 0cm 0cm}, clip, scale=0.35]{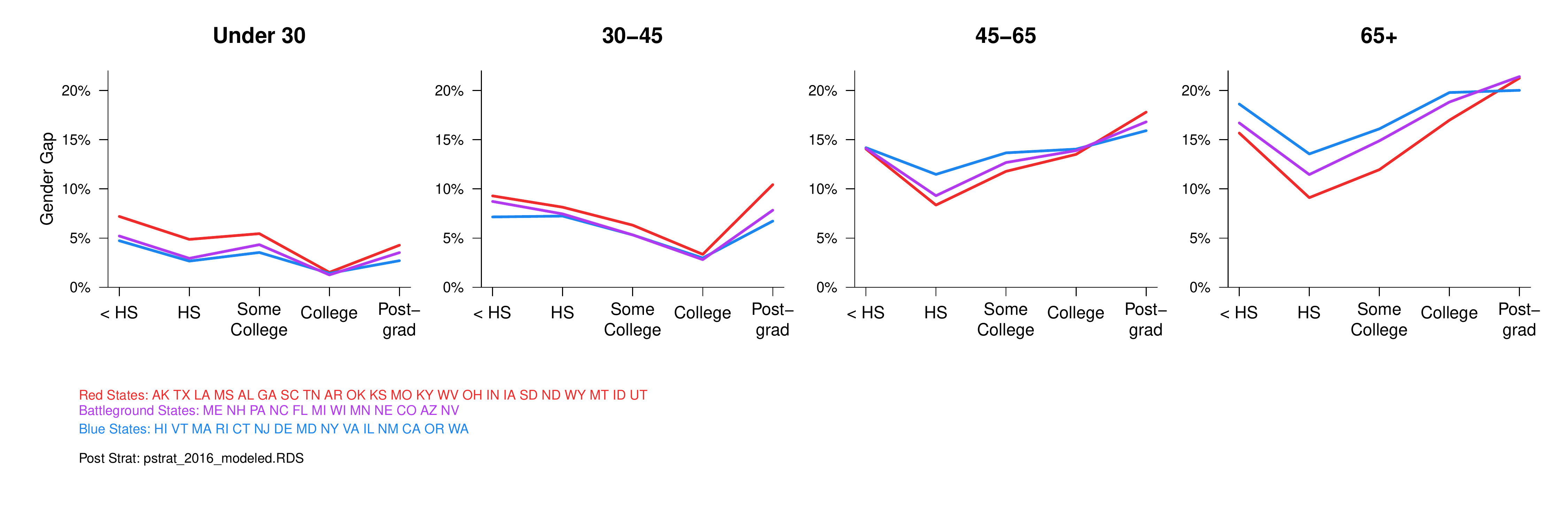}
\footnotesize
\emph{Notes:} The gender gap is evaluated as men's probability of voting for Trump minus women's probability for of voting for Trump for various education levels. Larger values indicate a greater divergence in vote preference among women and men. Interactions exist between age and education conditional on gender. Overall, the gender gap increases with age. Among voters under 45 the gender gap is lowest for those with a college education, and among voters 45 years or older the gender gap is lowest for those with a high school education. \\
(Using \emph{Model 1}.)
\end{minipage}
\end{center}
\end{figure}
\begin{figure}[H]\caption[]{Gender Gap by Education for each Age Category - 2012 Election}
\begin{center}
\begin{minipage}{0.95\linewidth}
\includegraphics[trim={0cm 2.5cm 0cm 0cm}, clip, scale=0.35]{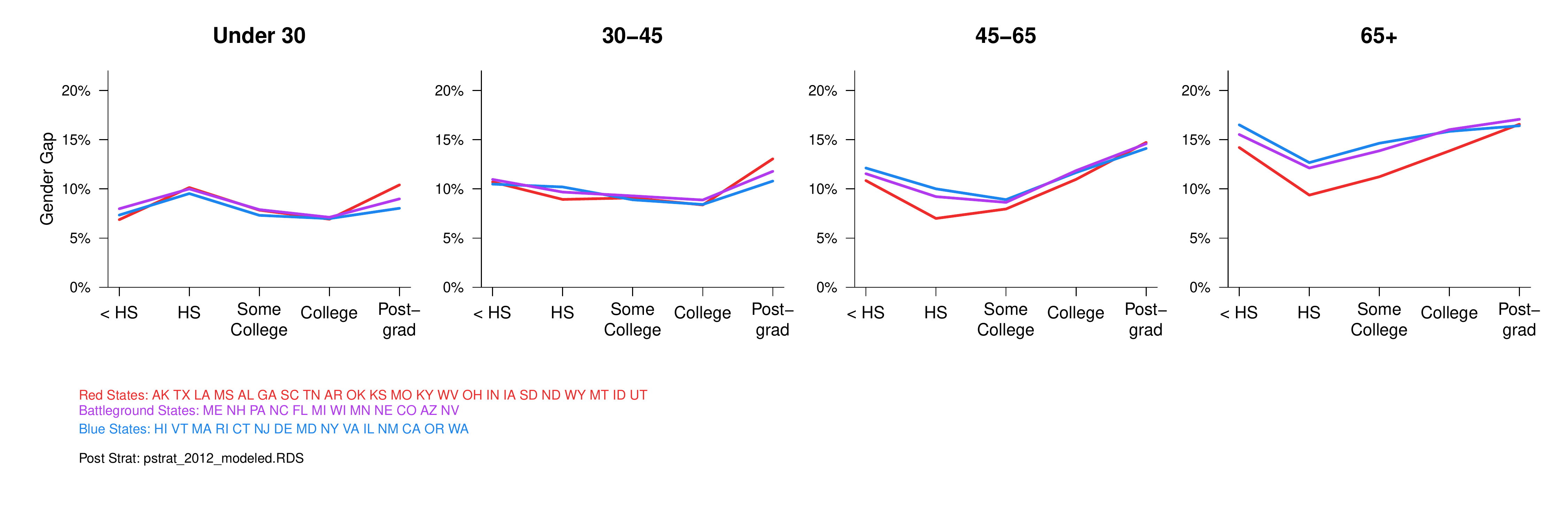}
\footnotesize
\emph{Notes:} The gender gap is evaluated as men's probability of voting for Romney minus women's probability for of voting for Romney for various education levels. Larger values indicate a greater divergence in vote preference among women and men. Interactions exist between age and education conditional on gender. \\
(Using \emph{Model 1} with 2012 election results/turnout data.)
\end{minipage}
\end{center}
\end{figure}
\begin{figure}[H]\caption[]{Gender Gap by Education (Men minus Women)}
\begin{minipage}{1\linewidth}
\includegraphics[trim={2.5cm 5cm 4cm 3.5cm}, clip, scale=0.23]{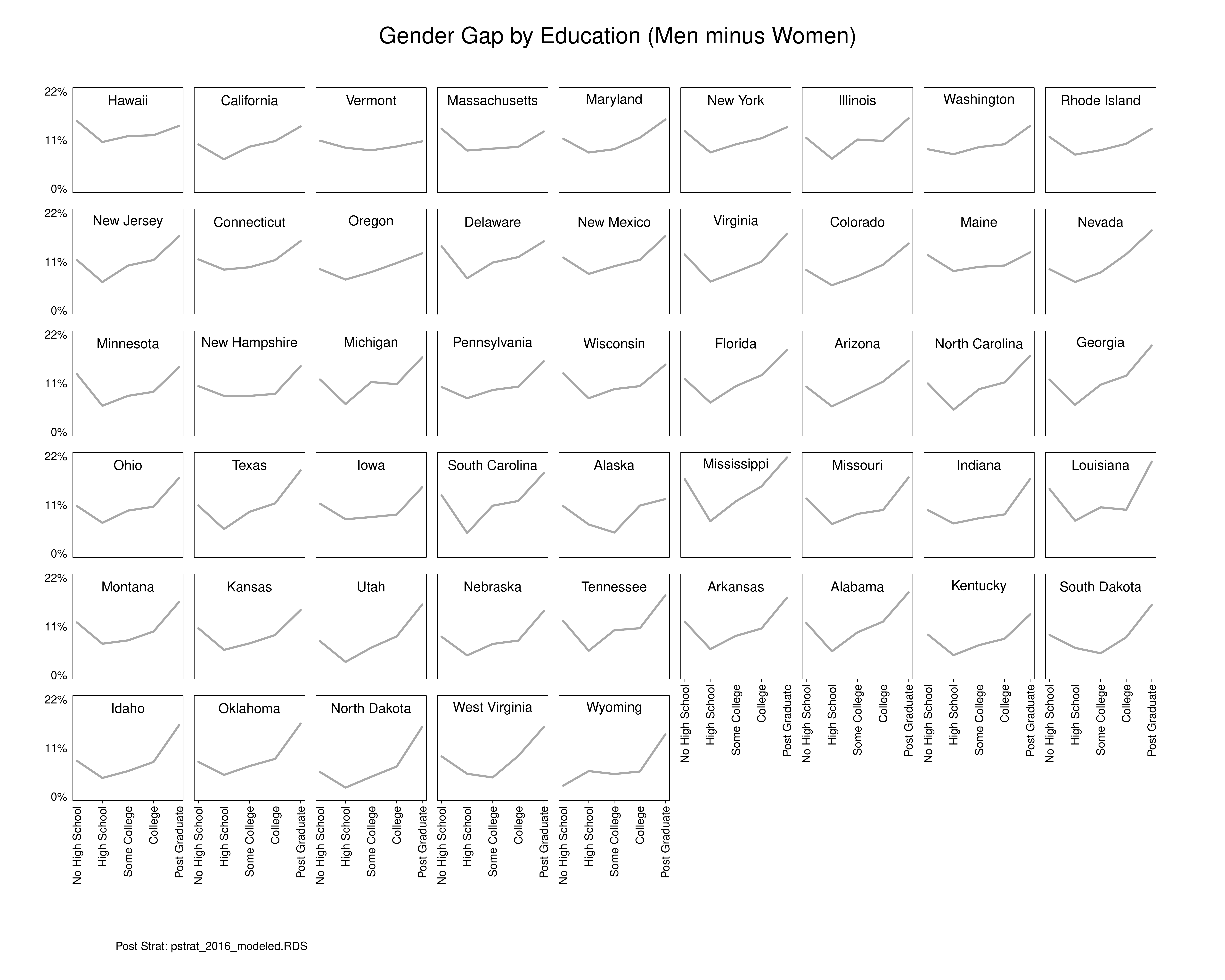}
\footnotesize
\emph{Notes:} The state-level gender gap is evaluated as men's probability of voting for Trump minus women's probability for of voting for Trump for various education levels. Larger values indicate a greater divergence in vote preference among women and men. In most states, voters with a high school education level tend to have the lowest gender gap and voters with a post graduate education level tend to have the highest gender gap. \\
(Using \emph{Model 1}.)
\end{minipage}
\end{figure}
\begin{figure}[H]\caption[]{Gender Gap by Age (Men minus Women)}
\begin{center}
\begin{minipage}{1\linewidth}
\includegraphics[trim={0cm 7cm 0cm 5cm}, clip, scale=0.23]{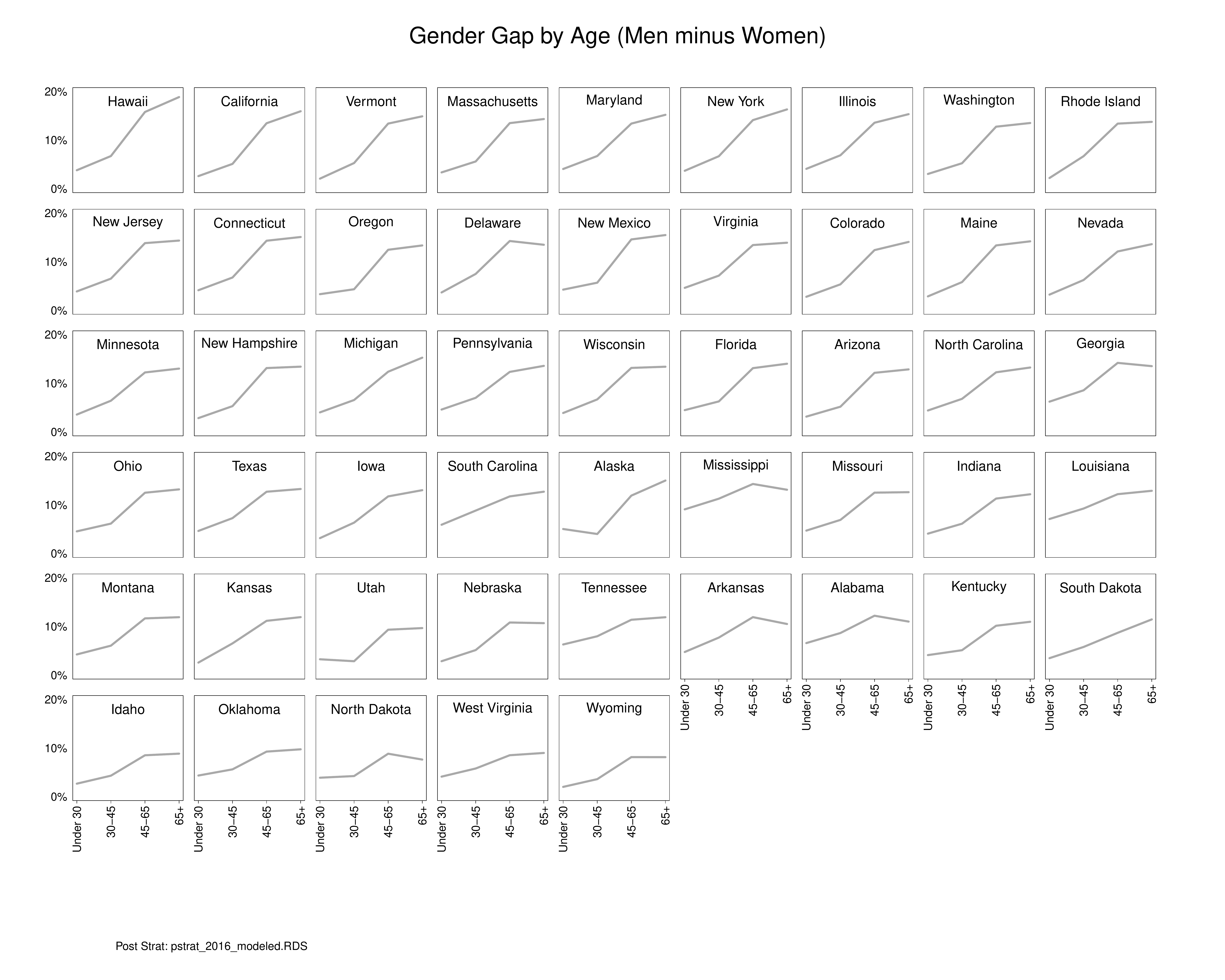}
\footnotesize
\emph{Notes:} The state-level gender gap is evaluated as men's probability of voting for Trump minus women's probability for of voting for Trump for various education levels. Larger values indicate a greater divergence in vote preference among women and men. The gender gap increases with age in most states, with larger variation in states that supported Clinton. \\
(Using \emph{Model 1}.)
\end{minipage}
\end{center}
\end{figure}

 \subsubsection{Vote by education}

\begin{figure}[H]\caption[]{Trump's Share of the Two-Party Vote by Education for each Age Category}
\begin{center}
\begin{minipage}{0.95\linewidth}
\includegraphics[trim={0cm 2.5cm 0cm 0cm}, clip, scale=0.35]{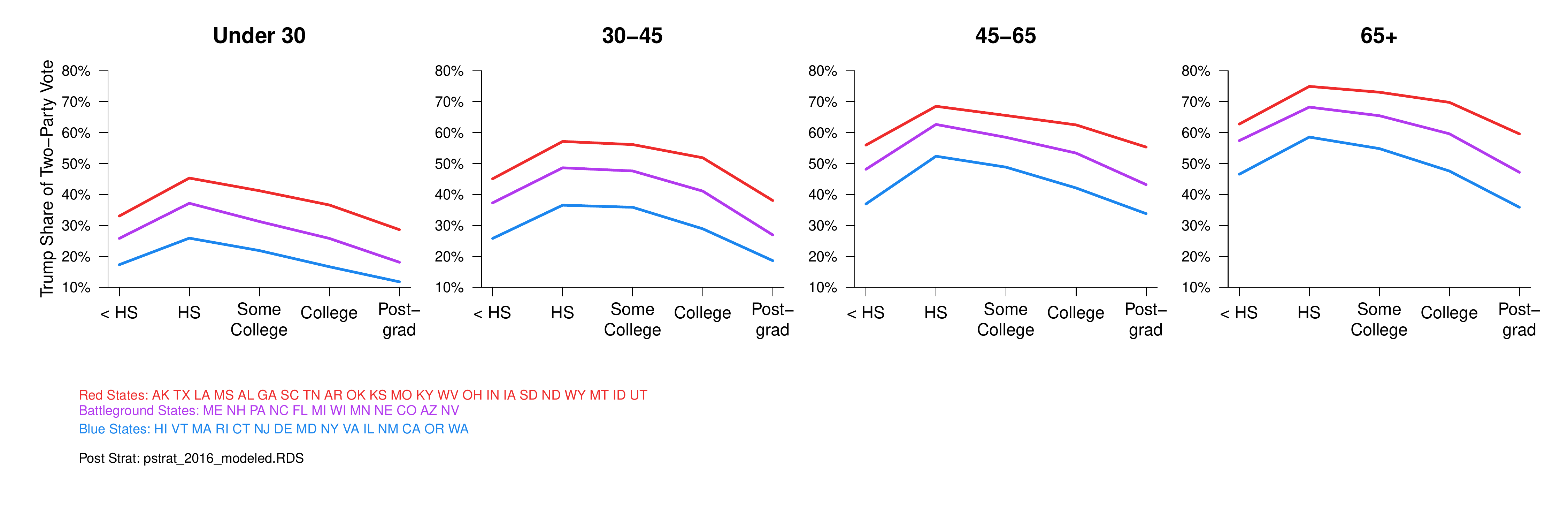}
\footnotesize
\emph{Notes:} Republican share of the two-party vote against various education levels. Overall, the Republican share increases with age. The strongest support came from voters with a high school education in each age category, with the exception of 30-45 year olds. \\
(Using \emph{Model 1}.)
\end{minipage}
\end{center}
\end{figure}
\begin{figure}[H]\caption[]{Romney's Share of the Two-Party Vote by Education for each Age Category - 2012 Election}
\begin{center}
\begin{minipage}{0.95\linewidth}
\includegraphics[trim={0cm 2.5cm 0cm 0cm}, clip, scale=0.35]{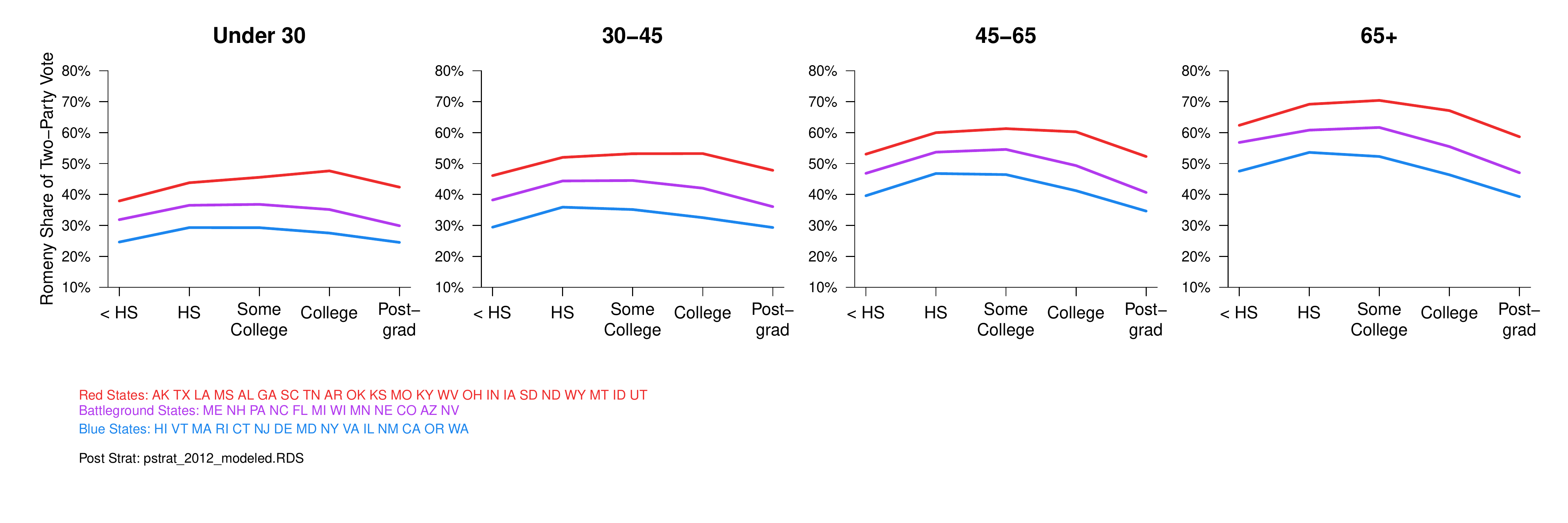}
\footnotesize
\emph{Notes:} Republican share of the two-party vote against various education levels. Overall, the Republican share increases with age. \\
(Using \emph{Model 1} with 2012 election results/turnout data.)
\end{minipage}
\end{center}
\end{figure}

\subsubsection{Vote by income, age, education, and ethnicity}

\begin{figure}[H]\caption[]{Trump's Share of the Two-Party Vote by Income and Education}
\begin{center}
\begin{minipage}{0.85\linewidth}
\includegraphics[trim={0cm 0.5cm 0cm 2cm}, clip, scale=0.35]{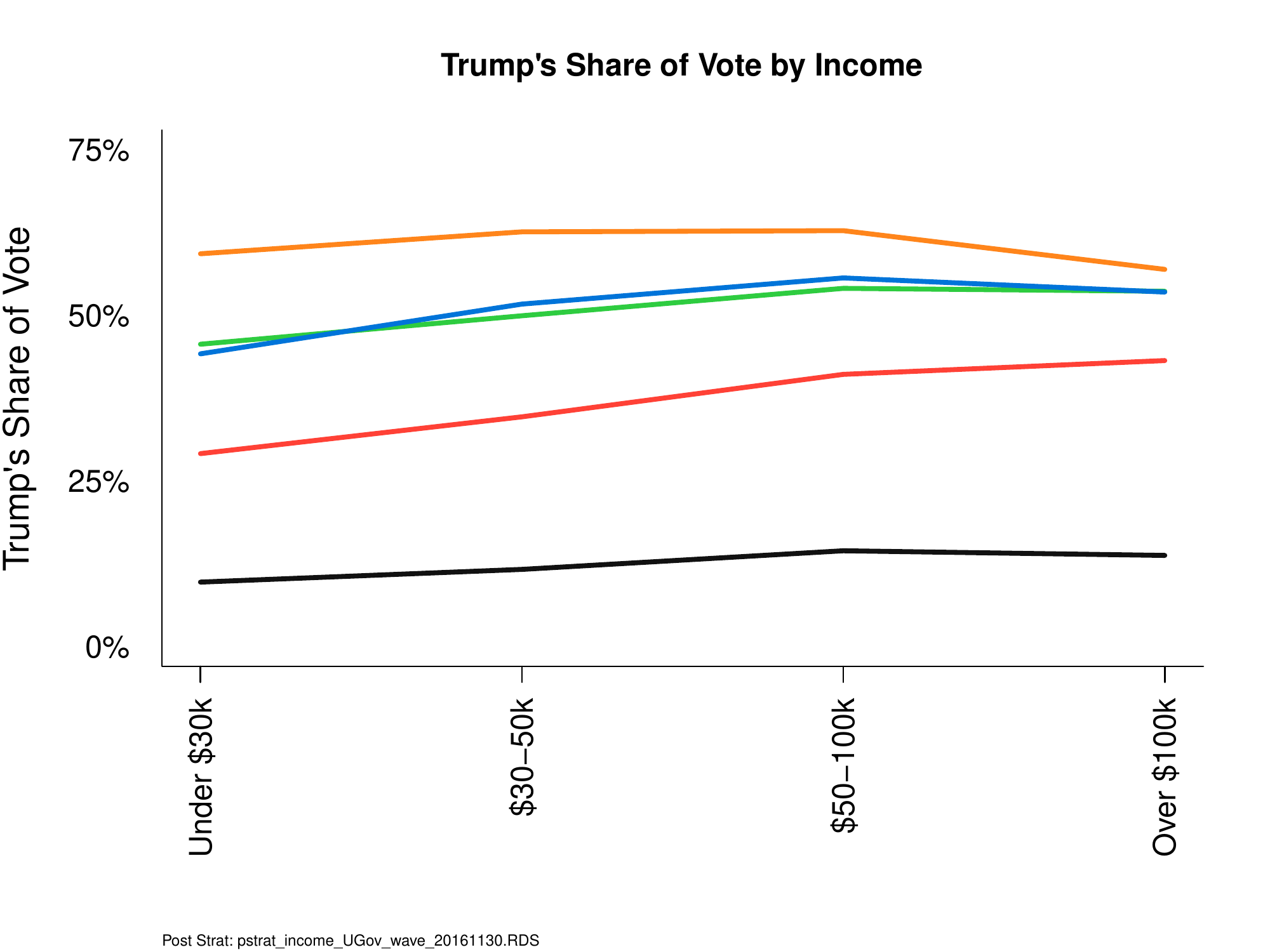}
\includegraphics[trim={0.7cm 0.5cm 0cm 2cm}, clip, scale=0.35]{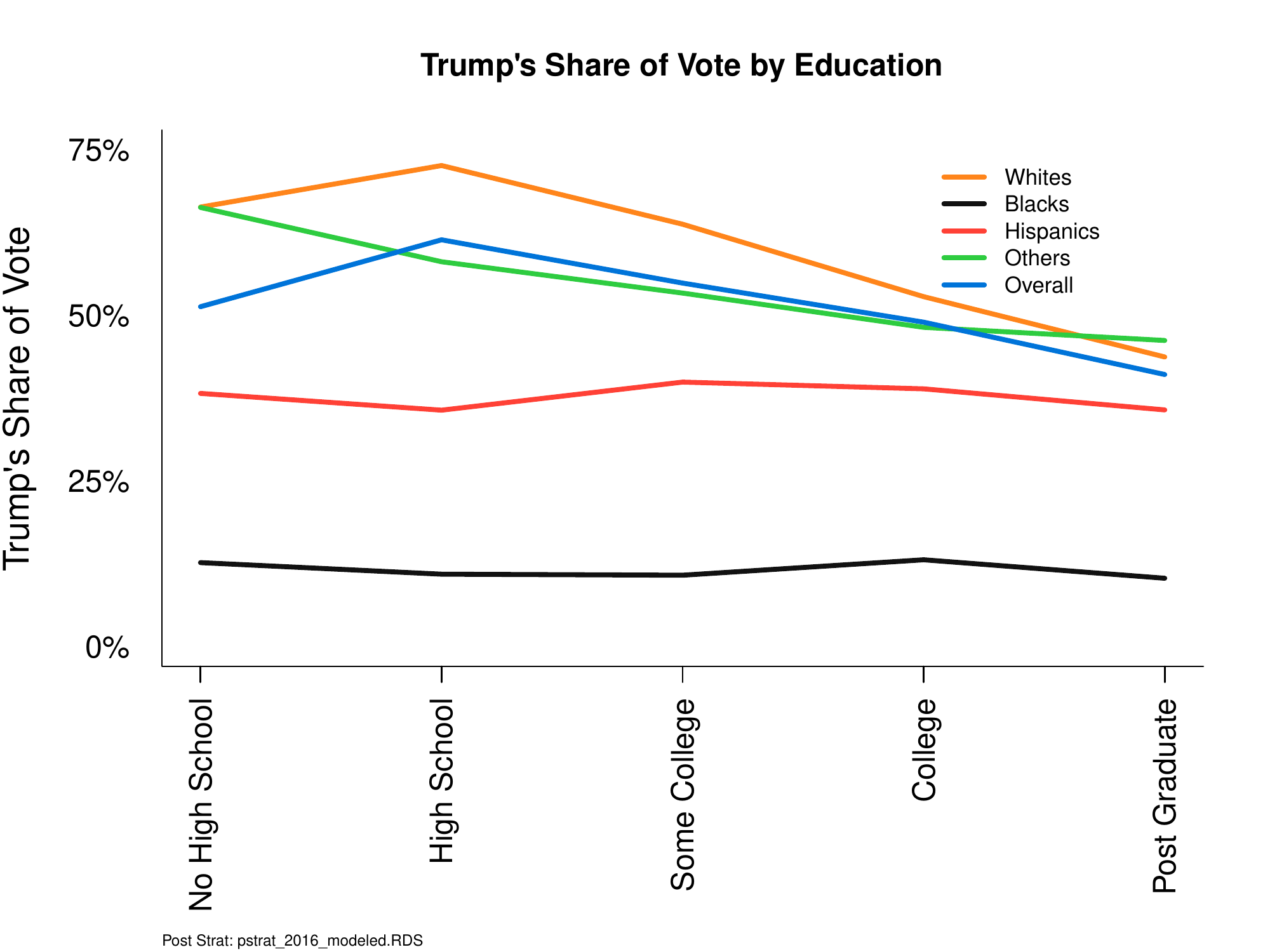}
\footnotesize
\emph{Notes:} Republican share of the two-party vote for Whites (orange), Blacks (black), Hispanics (red), other ethnicities (green), and overall (blue). Trump's share of the vote is highest among white voters with a high school education level. \\
(Using \emph{Model 2} (left) and \emph{Model 1} (right).)
\end{minipage}
\end{center}
\end{figure}
\begin{figure}[H]\caption[]{Trump's Share of the Two-Party Vote by Education, Ethnicity, and State}
\begin{center}
\begin{minipage}{1\linewidth}
\includegraphics[trim={0cm 5cm 0cm 5cm}, clip, scale=0.23]{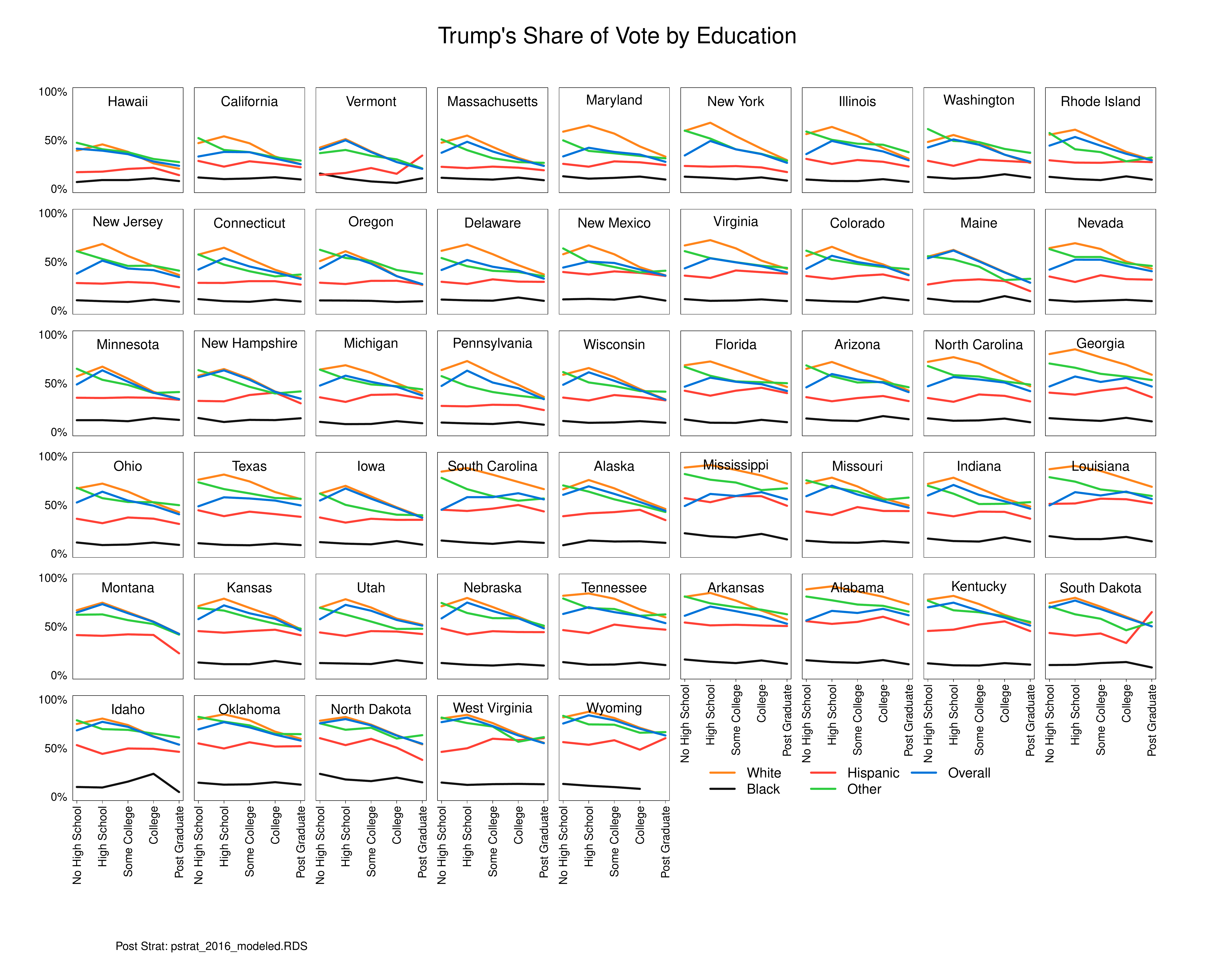}
\footnotesize
\emph{Notes:} State-level Republican share of the two-party vote for Whites (orange), Blacks (black), Hispanics (red), other ethnicities (green), and overall (blue). In most states white voters with high school education have the greatest support for Trump and those with post graduate education have the lowest support for Trump.  \\
(Using \emph{Model 1}.)
\end{minipage}
\end{center}
\end{figure}
\begin{figure}[H]\caption[]{Romney's Share of the Two-Party Vote by Education, Ethnicity, and State - 2012 Election}
\begin{center}
\begin{minipage}{1\linewidth}
\includegraphics[trim={0cm 5cm 0cm 5cm}, clip, scale=0.23]{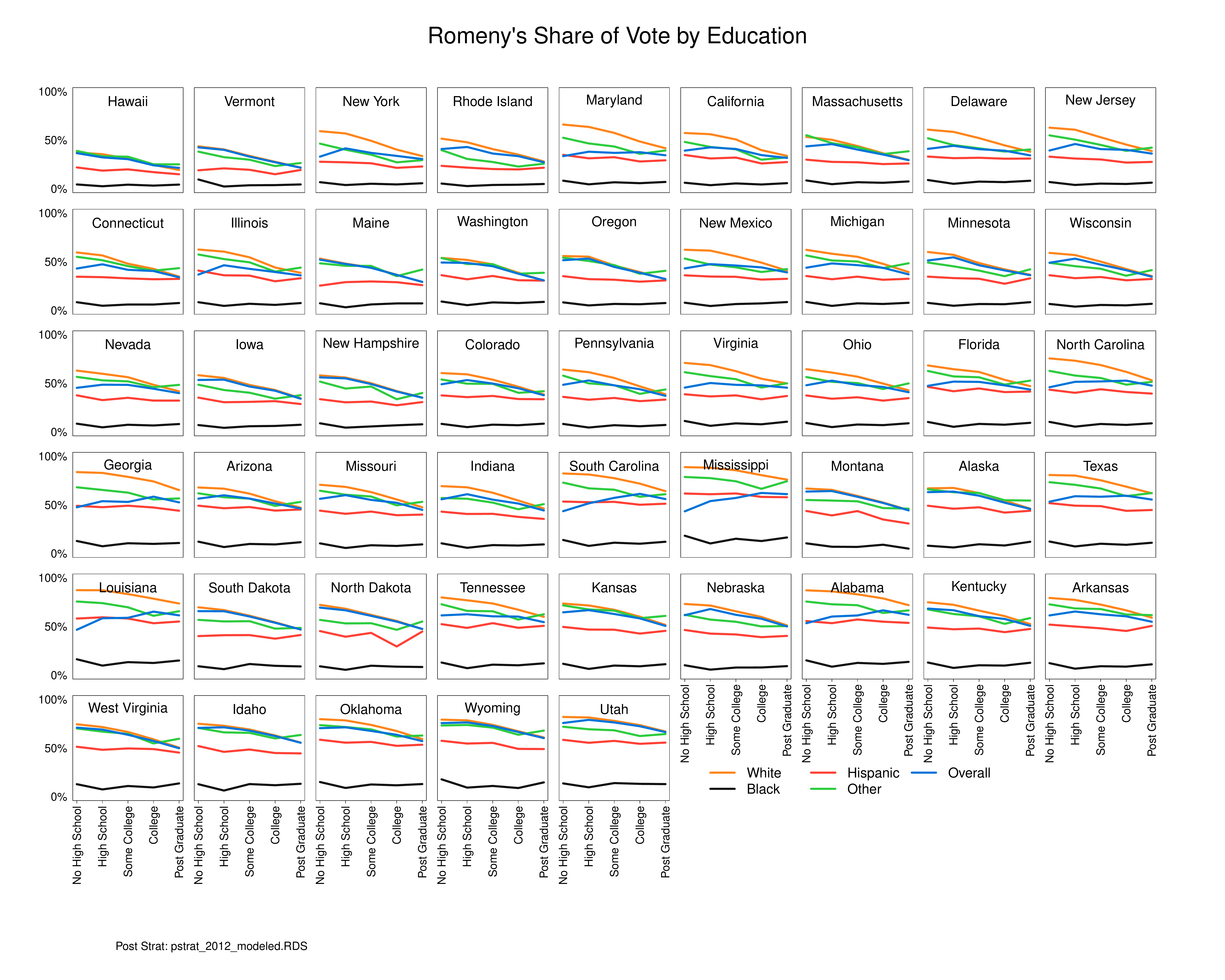}
\footnotesize
\emph{Notes:} State-level Republican share of the two-party vote for Whites (orange), Blacks (black), Hispanics (red), other ethnicities (green), and overall (blue).  \\
(Using \emph{Model 1} with 2012 election results/turnout data.)
\end{minipage}
\end{center}
\end{figure}
\begin{figure}[H]\caption[]{Trump's Share of the Two-Party Vote by Age, Ethnicity, and State}
\begin{center}
\begin{minipage}{1\linewidth}
\includegraphics[trim={0cm 5cm 0cm 5cm}, clip, scale=0.23]{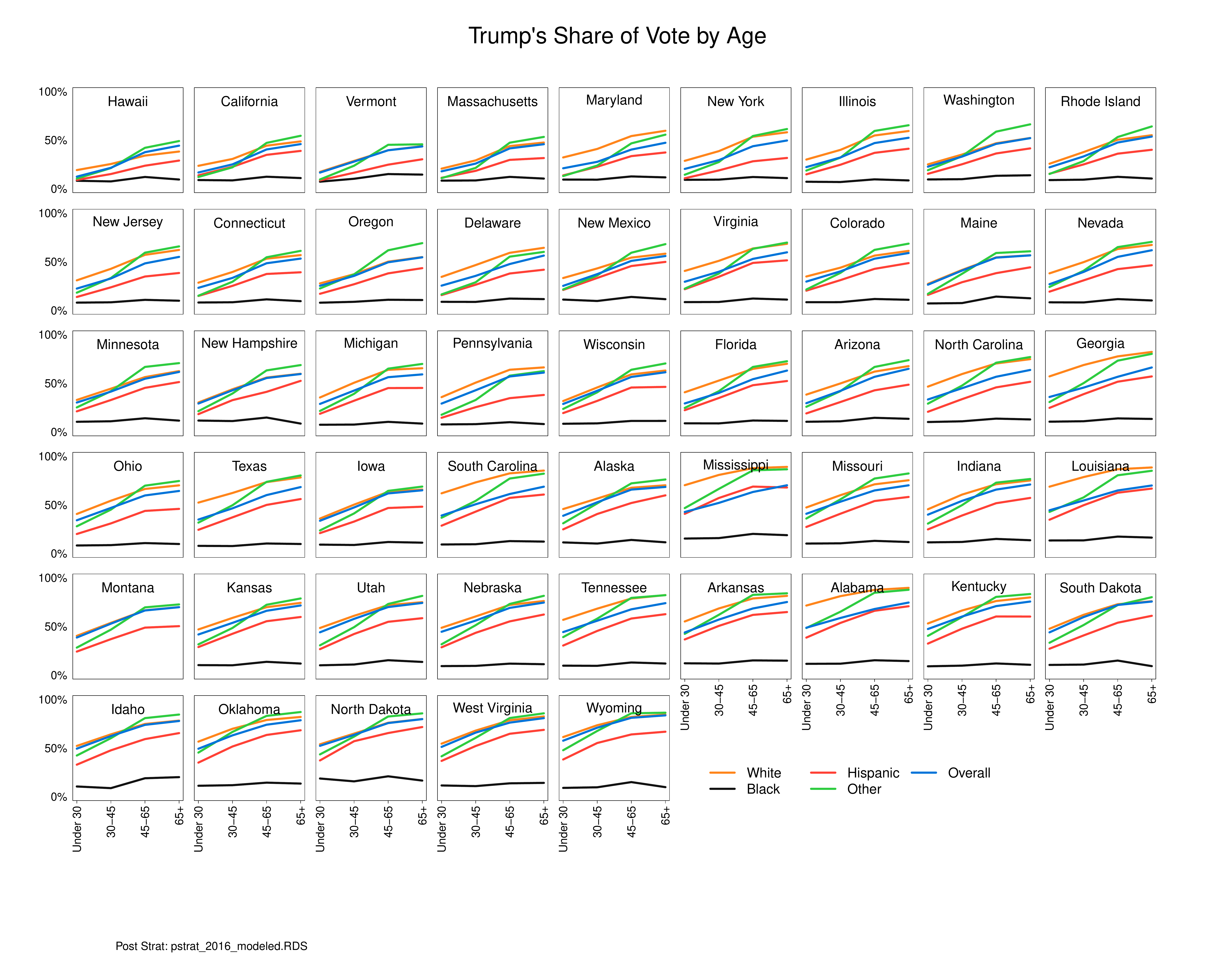}
\footnotesize
\emph{Notes:} State-level Republican share of the two-party vote for Whites (orange), Blacks (black), Hispanics (red), other ethnicities (green), and overall (blue). Support for Trump increases with age. Support among Whites is consistently the strongest followed by support among other races and Hispanics.  \\
(Using \emph{Model 1}.)
\end{minipage}
\end{center}
\end{figure}
\begin{figure}[H]\caption[]{Romney's Share of the Two-Party Vote by Age, Ethnicity, and State - 2012 Election}
\begin{center}
\begin{minipage}{1\linewidth}
\includegraphics[trim={0cm 5cm 0cm 5cm}, clip, scale=0.23]{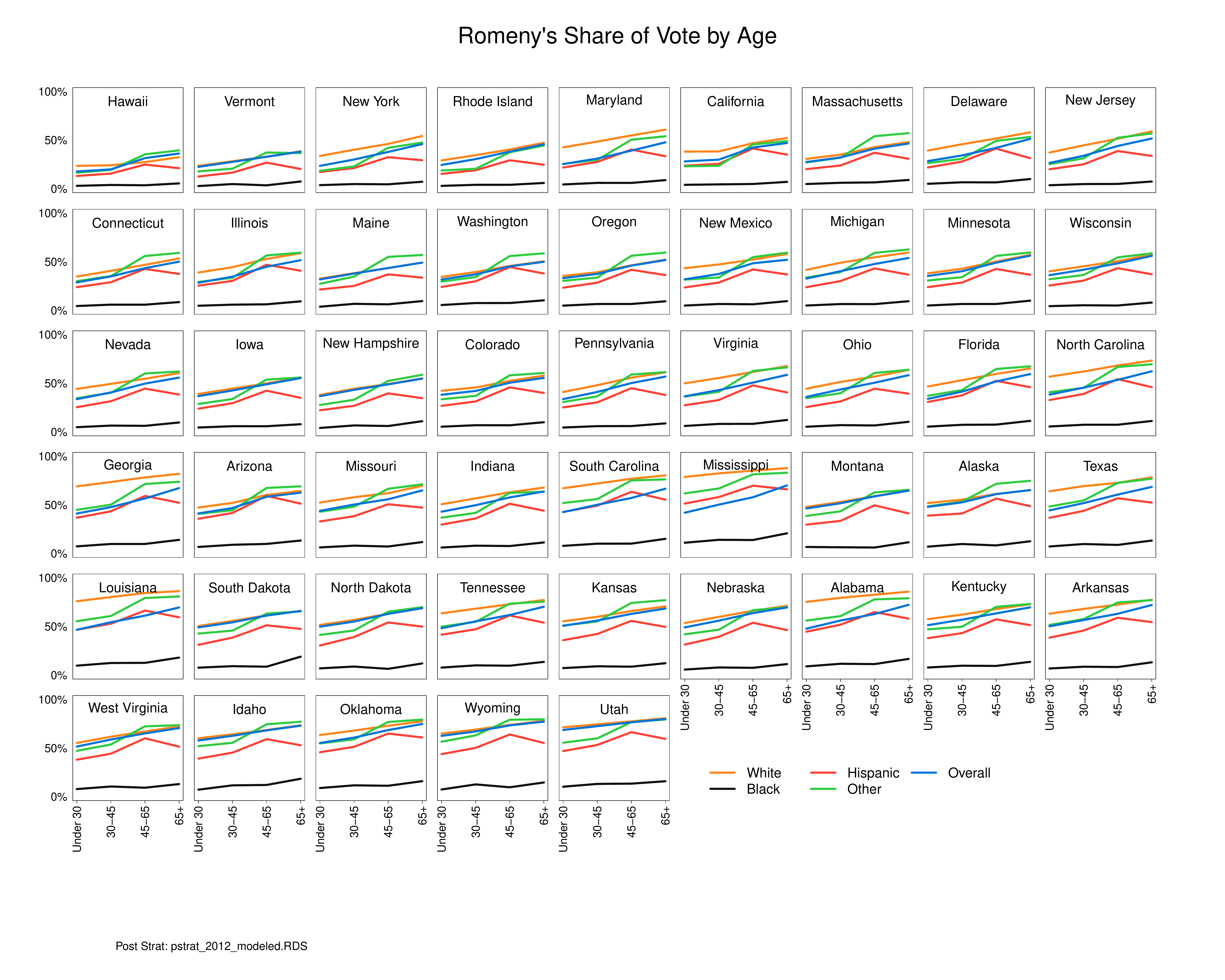}
\footnotesize
\emph{Notes:} State-level Republican share of the two-party vote for Whites (orange), Blacks (black), Hispanics (red), other ethnicities (green), and overall (blue). Support for Trump increases with age. \\
(Using \emph{Model 1} with 2012 election results/turnout data.)
\end{minipage}
\end{center}
\end{figure}

\subsubsection{Voter turnout}

\begin{figure}[H]\caption[]{Voter Turnout by Education, Ethnicity and State}
\begin{center}
\begin{minipage}{1\linewidth}
\includegraphics[trim={0cm 5cm 0cm 5cm}, clip, scale=0.23]{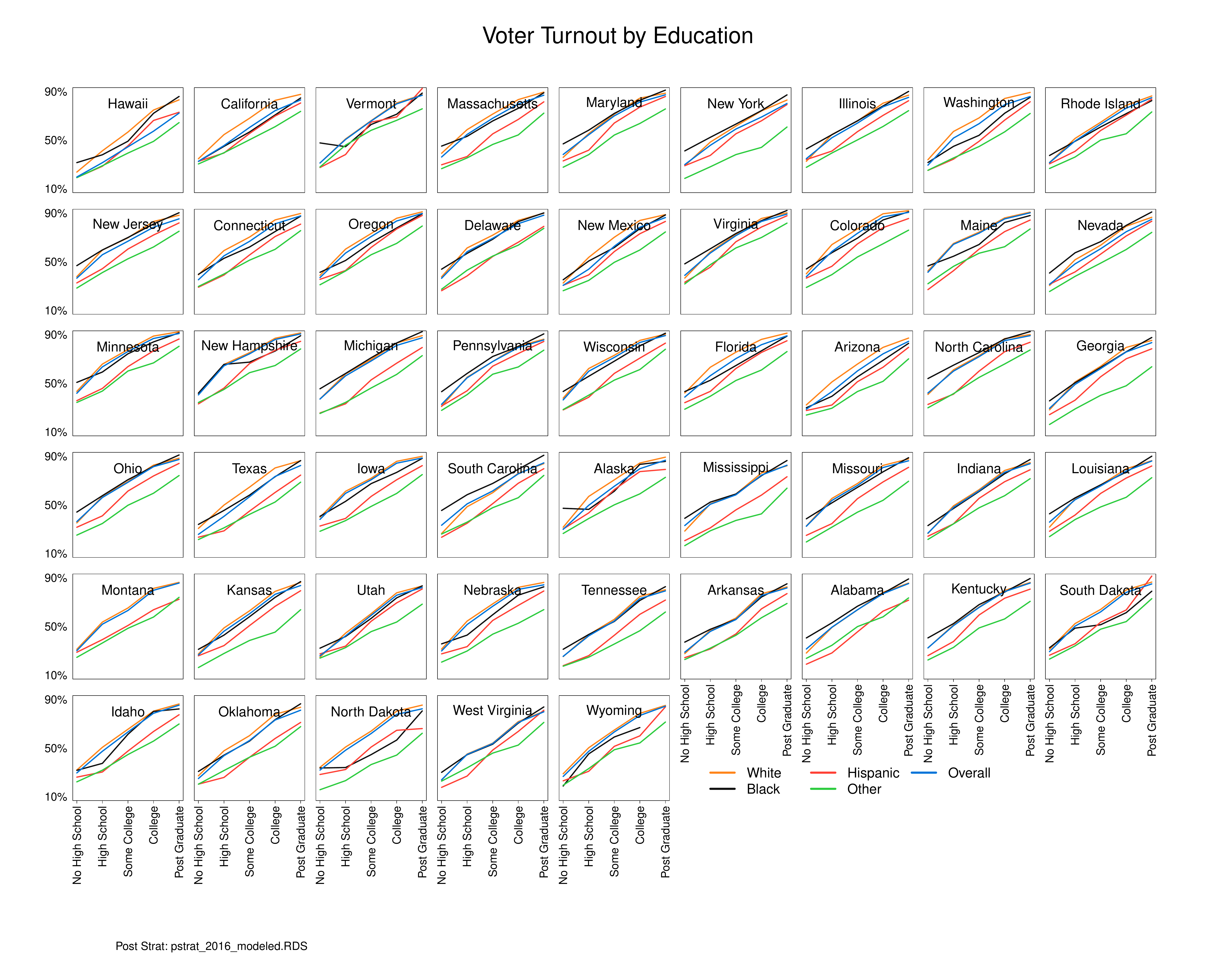}
\footnotesize
\emph{Notes:} Voter turnout for Whites (orange), Blacks (black), Hispanics (red), other ethnicities (green), and overall (blue). Voter turnout increases with education. There is not much variation across states. Within states Hispanics typically experienced low voter turnout compared to Whites and Blacks. \\
(Using \emph{Model 1}.)
\end{minipage}
\end{center}
\end{figure}

\begin{figure}[H]\caption[]{Voter Turnout by Education, Ethnicity and State - 2012 Election}
\begin{center}
\begin{minipage}{1\linewidth}
\includegraphics[trim={0cm 5cm 0cm 5cm}, clip, scale=0.23]{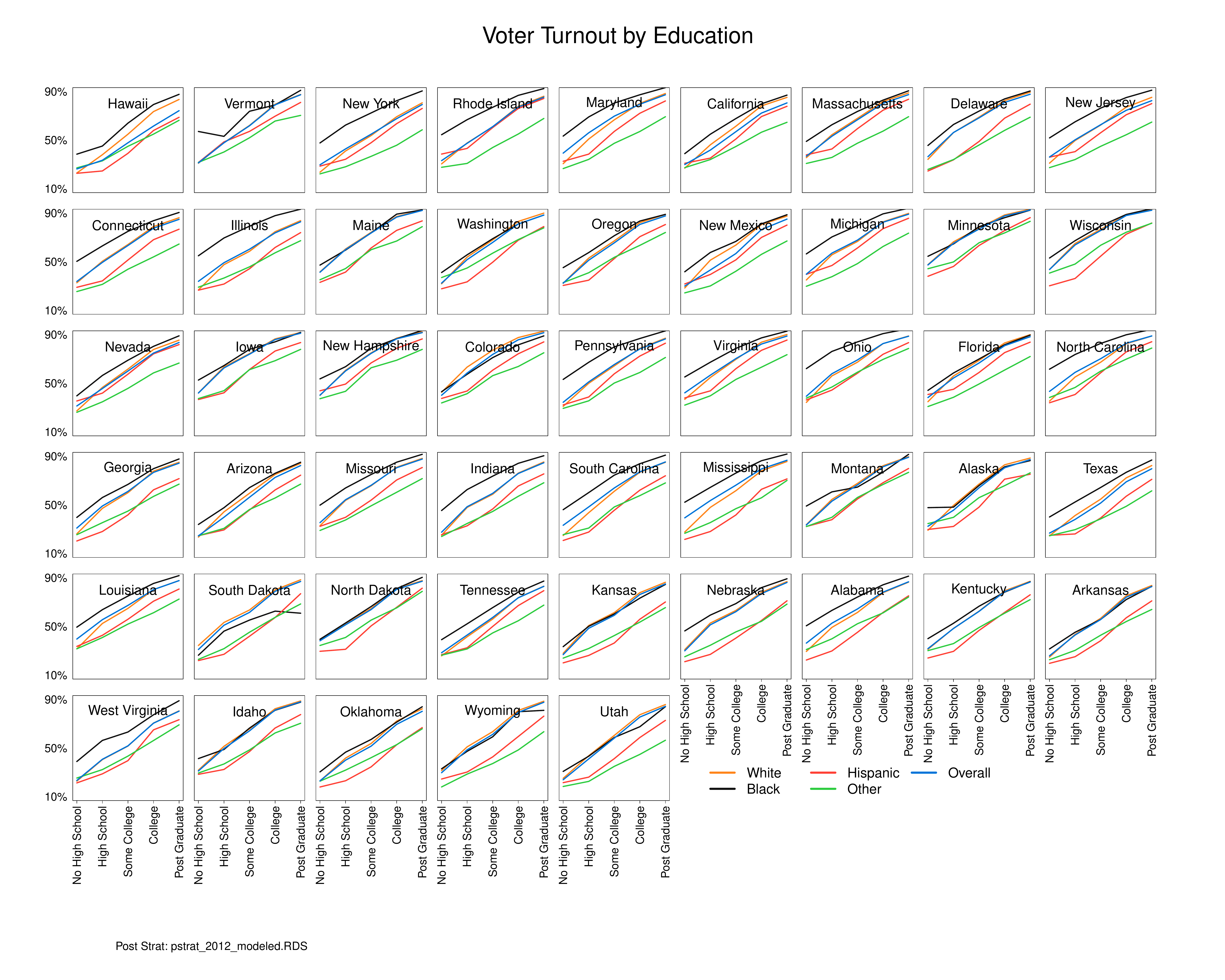}
\footnotesize
\emph{Notes:} Voter turnout for Whites (orange), Blacks (black), Hispanics (red), other ethnicities (green), and overall (blue).\\
(Using \emph{Model 1} with 2012 election results/turnout data.)
\end{minipage}
\end{center}
\end{figure}
\begin{figure}[H]\caption[]{Voter Turnout by Age, Ethnicity and State}
\begin{center}
\begin{minipage}{1\linewidth}
\includegraphics[trim={0cm 5cm 0cm 5cm}, clip, scale=0.23]{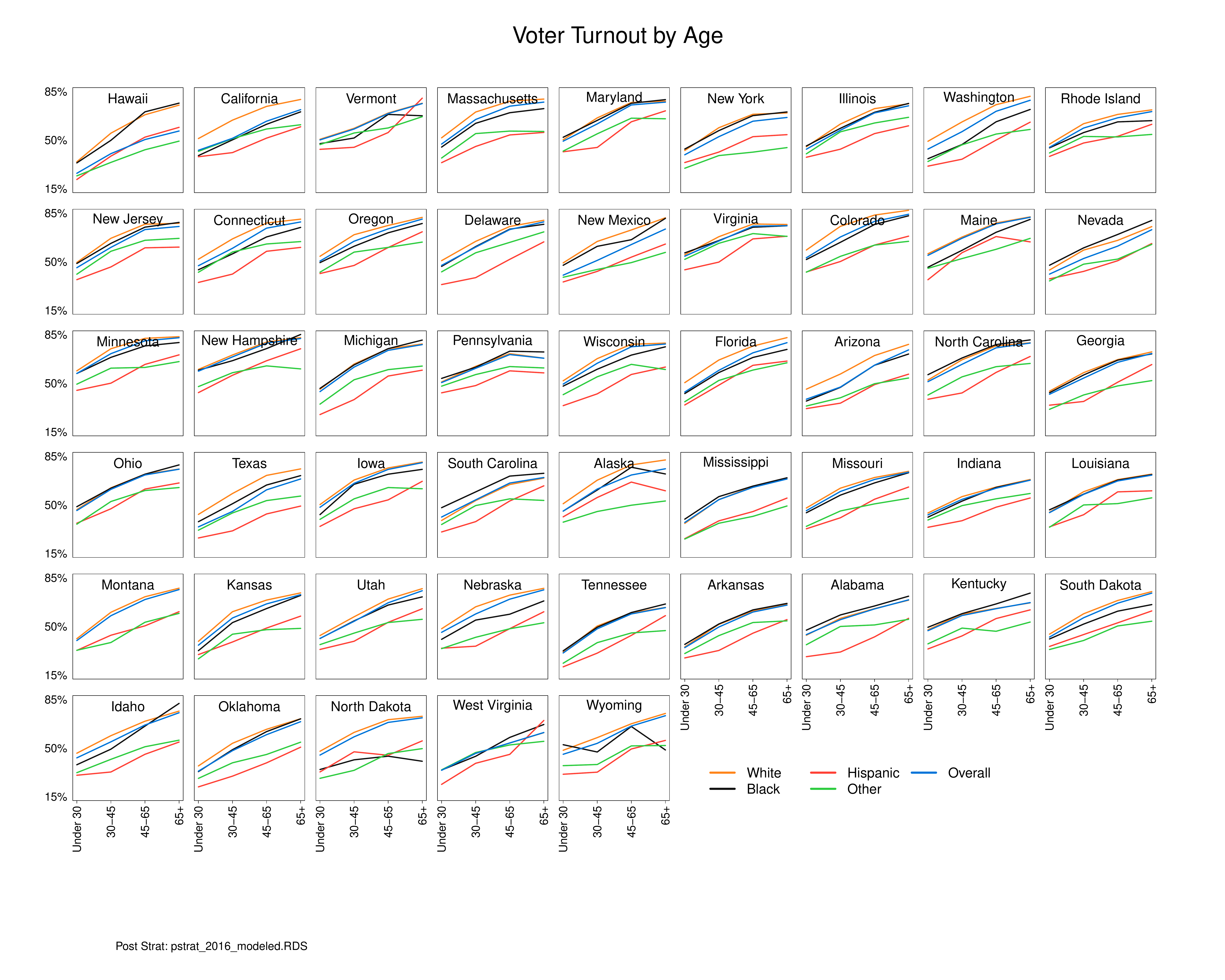}
\footnotesize
\emph{Notes:} Voter turnout for Whites (orange), Blacks (black), Hispanics (red), other ethnicities (green), and overall (blue). Voter turnout increases with age. There is low voter turnout among Hispanics across age levels compared to Whites and Blacks.\\
(Using \emph{Model 1}.)
\end{minipage}
\end{center}
\end{figure}
\begin{figure}[H]\caption[]{Voter Turnout by Age, Ethnicity and State - 2012 Election}
\begin{center}
\begin{minipage}{1\linewidth}
\includegraphics[trim={0cm 5cm 0cm 5cm}, clip, scale=0.23]{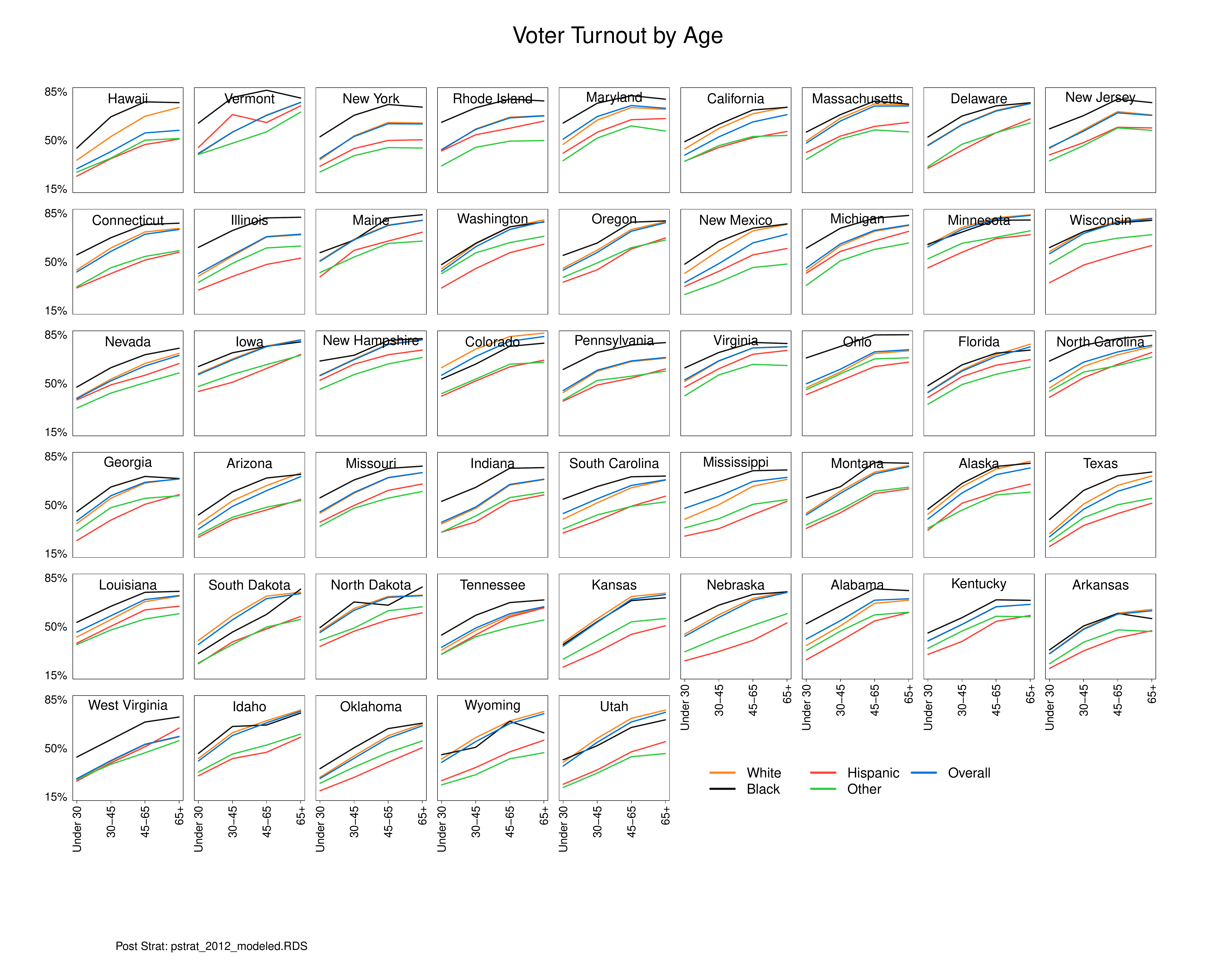}
\footnotesize
\emph{Notes:} Voter turnout for Whites (orange), Blacks (black), Hispanics (red), other ethnicities (green), and overall (blue).\\
(Using \emph{Model 1} with 2012 election results/turnout data.)
\end{minipage}
\end{center}
\end{figure}
\begin{figure}[H]\caption[]{Voter Turnout by Education, Gender and State}
\begin{center}
\begin{minipage}{1\linewidth}
\includegraphics[trim={0cm 5cm 0cm 5cm}, clip, scale=0.23]{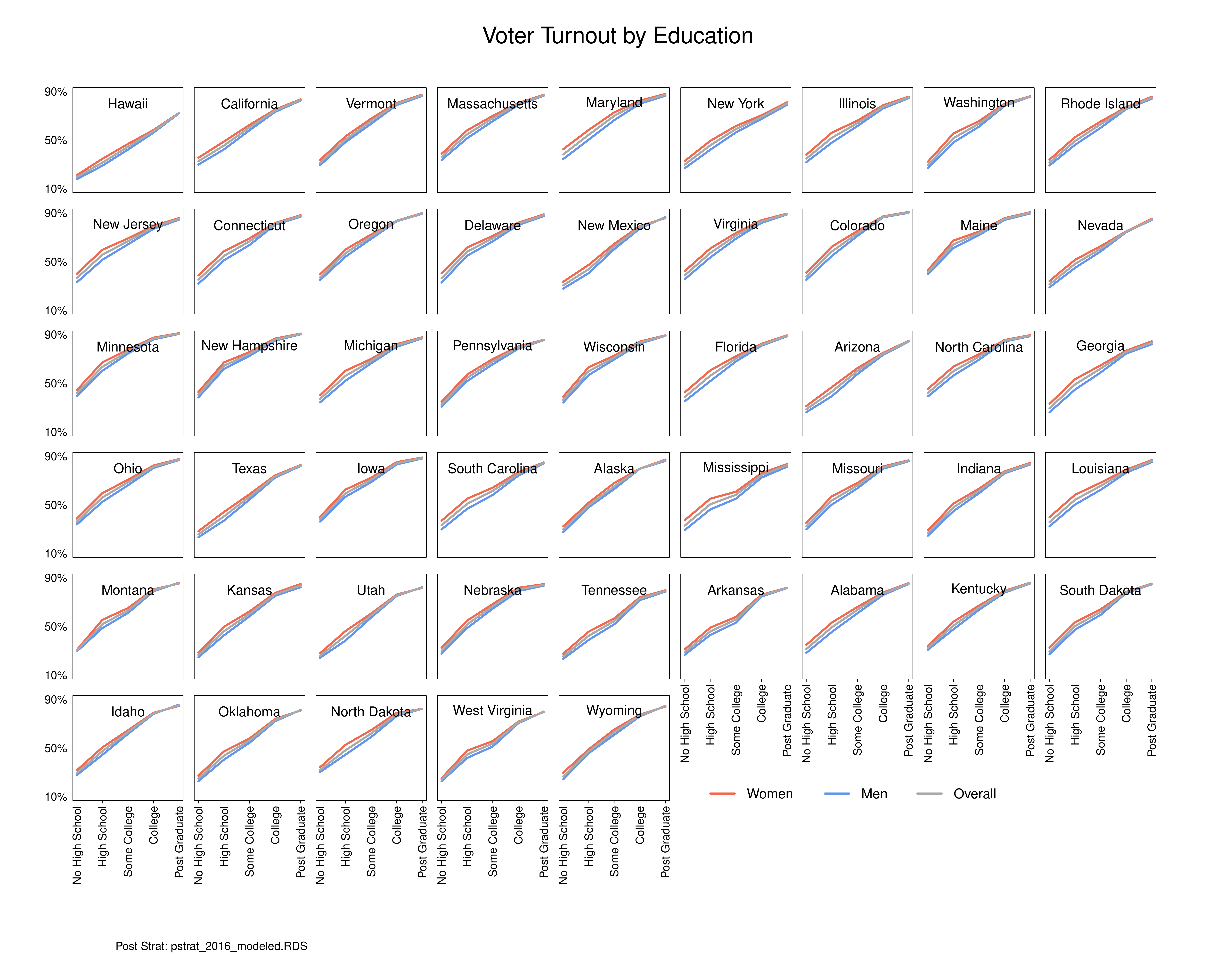}
\footnotesize
\emph{Notes:} Voter turnout for women (red), men (blue), and overall (grey). Voter turnout increases with education, with women experiencing a larger voter turnout compared to men. \\
(Using \emph{Model 1}.)
\end{minipage}
\end{center}
\end{figure}
\begin{figure}[H]\caption[]{Voter Turnout by Education, Gender and State - 2012 Election}
\begin{center}
\begin{minipage}{1\linewidth}
\includegraphics[trim={0cm 5cm 0cm 5cm}, clip, scale=0.23]{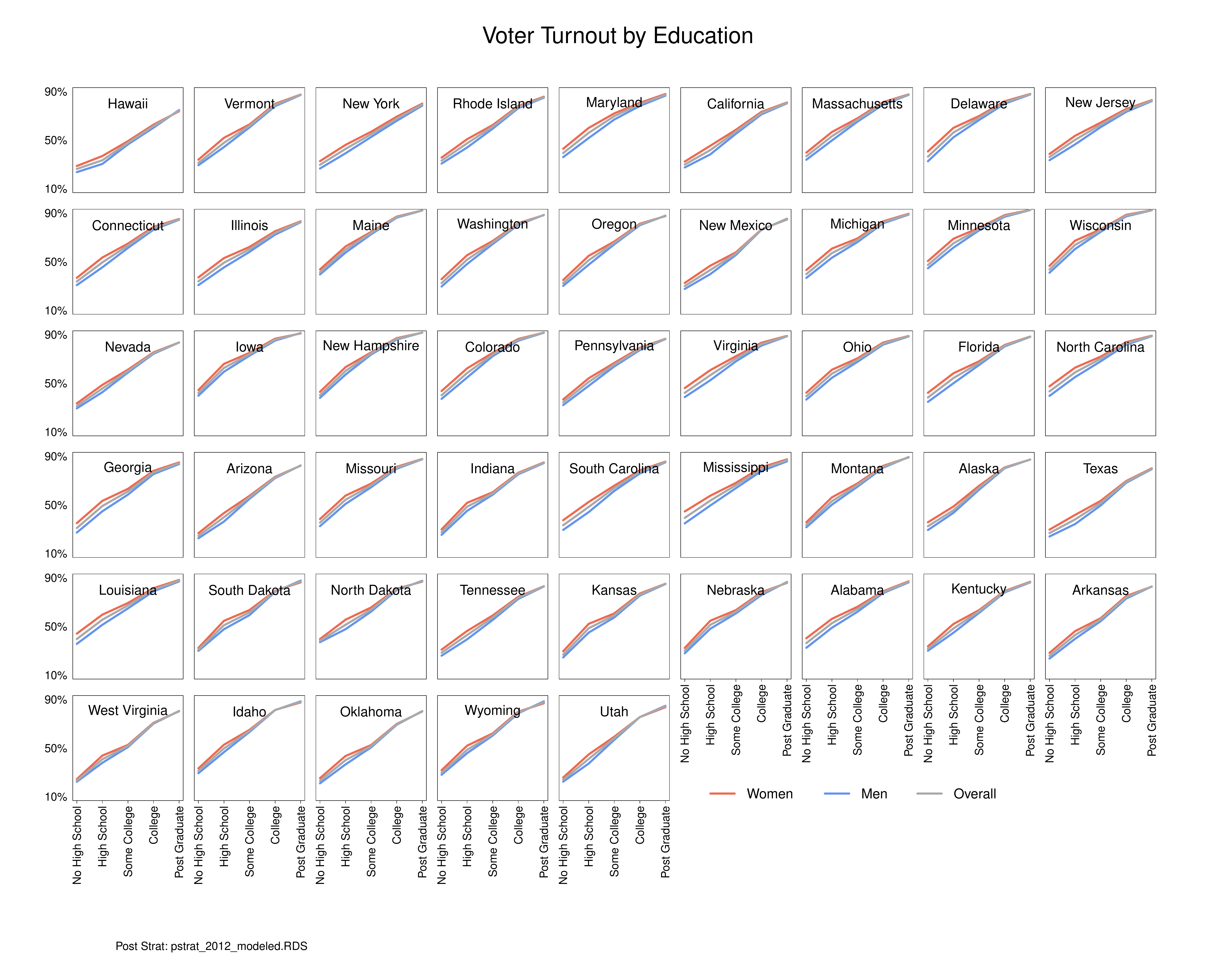}
\footnotesize
\emph{Notes:} Voter turnout for women (red), men (blue), and overall (grey). \\
(Using \emph{Model 1} with 2012 election results/turnout data.)
\end{minipage}
\end{center}
\end{figure}

\subsubsection{Maps of vote preference}

\begin{figure}[H]\caption[]{Gender Gap (Men minus Women)}
  \begin{minipage}{1\linewidth}
\includegraphics[trim={0cm 2cm 0cm 2cm}, clip, scale=0.4]{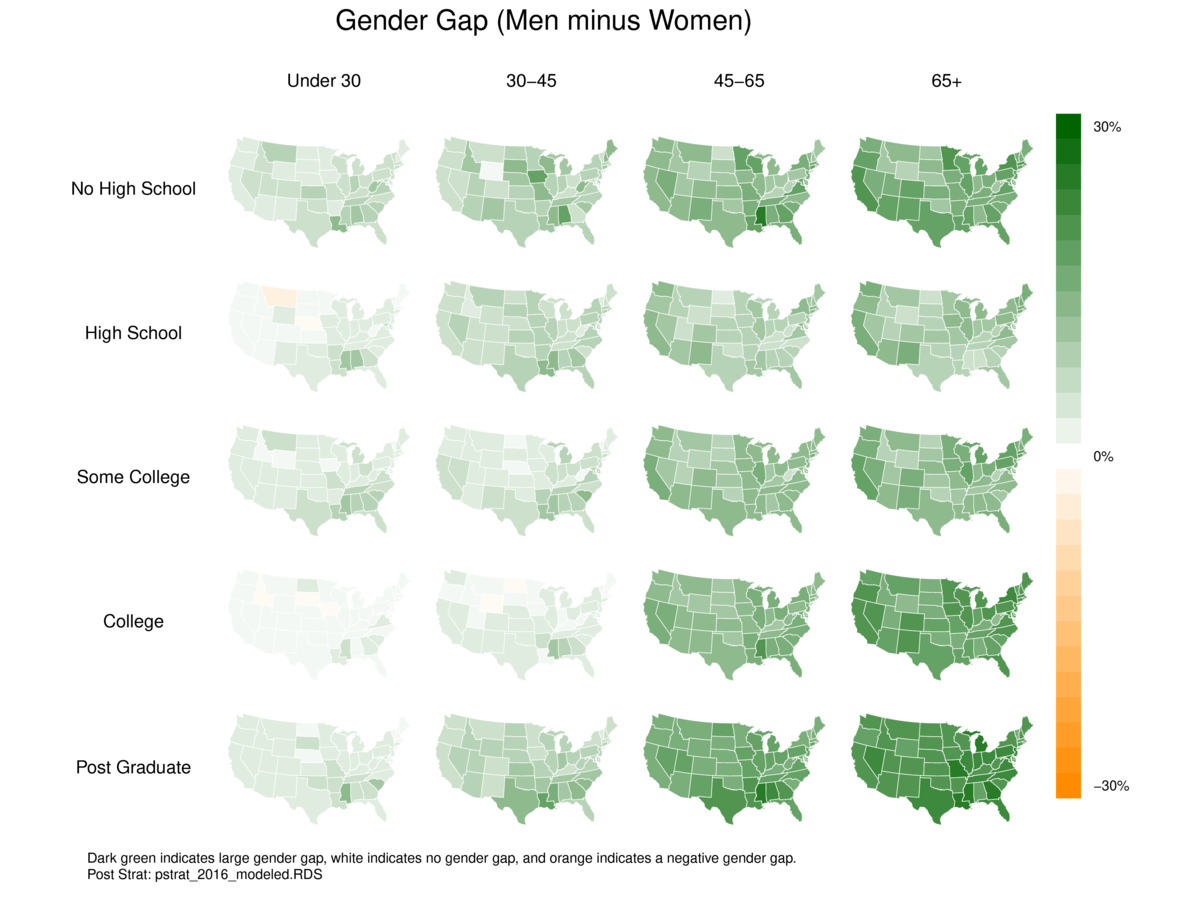}
\footnotesize
\emph{Notes:} State-level gender gap evaluated as men's probability of voting for Trump minus women's probability for of voting for Trump. Dark green/orange indicates a larger divergence in vote preference between men and women. The greatest divergence exists among older voters with post graduate education. The weakest support exists among young voters with a college education. \\
(Using \emph{Model 1}.)
  \end{minipage}
\end{figure}
\begin{figure}[H]\caption[]{Trump's Share of the Two-Party Vote by Age and Education}
\begin{minipage}{1\linewidth}
\includegraphics[trim={0cm 2cm 0cm 2cm}, clip, scale=0.4]{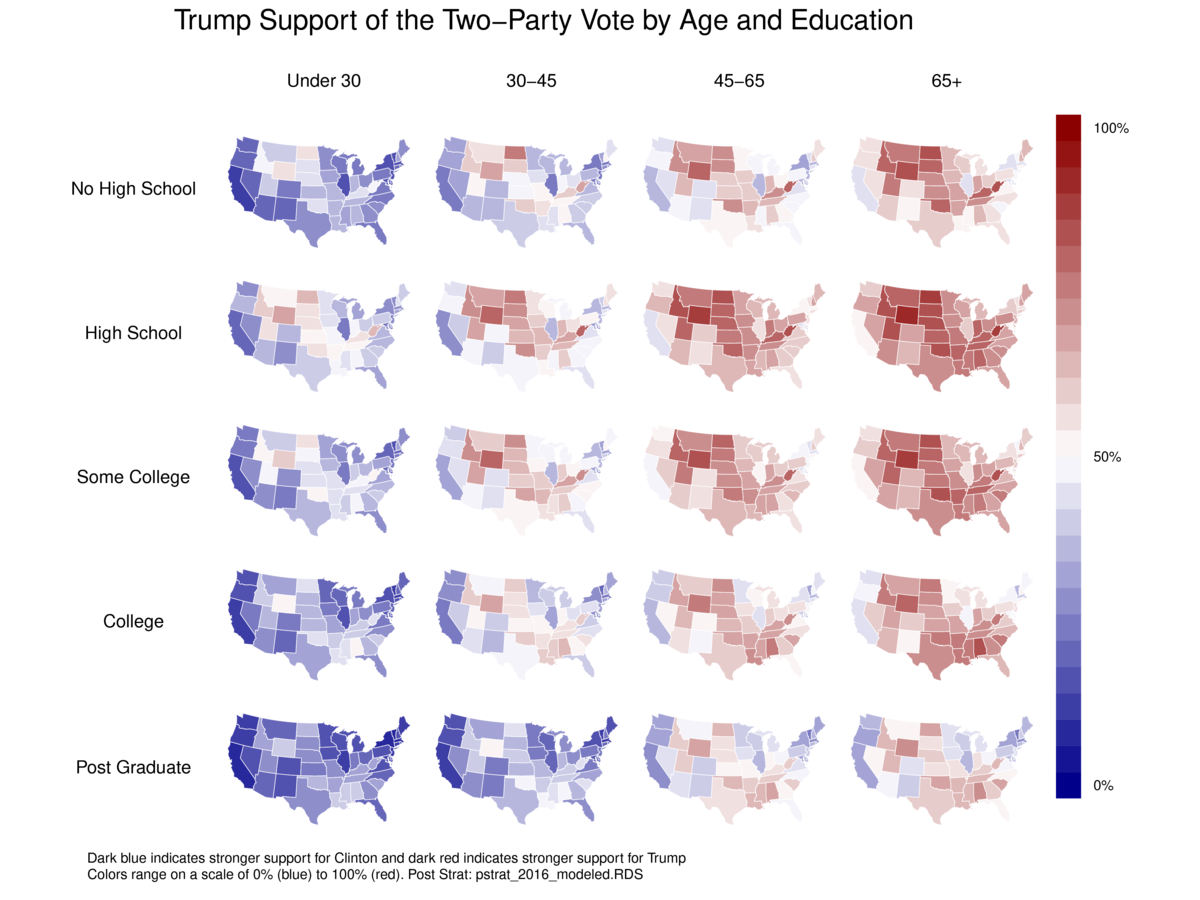}
\footnotesize
\emph{Notes:} State-level vote intention by education and age. Dark red indicates stronger support for Donald Trump and dark blue indicates stronger support for Hillary Clinton. Overall, older voters with lower education have stronger support for Trump and younger voters with higher levels of education have stronger support for Clinton. In each age bracket Trump has stronger support among voters with high school and some college education compared to voters with no high school education.\\
(Using \emph{Model 1}.)
\end{minipage}
\end{figure}
\begin{figure}[H]\caption[]{Trump's Share of the Two-Party Vote by Age and Education for Women}
\begin{minipage}{1\linewidth}
\includegraphics[trim={0cm 2cm 0cm 2cm}, clip, scale=0.4]{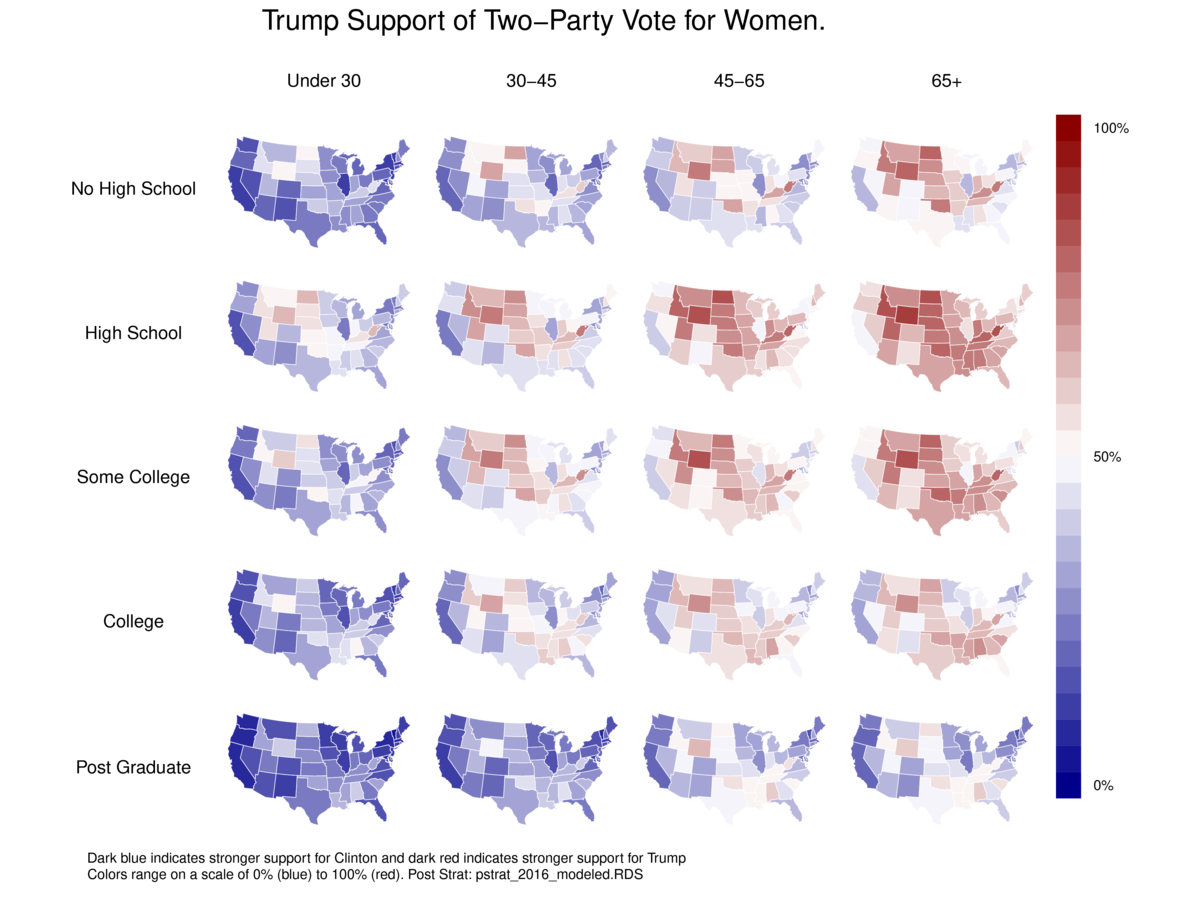}
\footnotesize
\emph{Notes:} State-level vote intention by education and age for women. Dark red indicates stronger support for Donald Trump and dark blue indicates stronger support for Hillary Clinton. Overall, older women have stronger support for Trump. Women with a post graduate education have stronger support for Clinton, and women with a high school education and some college education have stronger support for Trump.\\
(Using \emph{Model 1}.)
\end{minipage}
\end{figure}
\begin{figure}[H]\caption[]{Trump's Share of the Two-Party Vote by Age and Education for Men}
\begin{minipage}{1\linewidth}
\includegraphics[trim={0cm 2cm 0cm 2cm}, clip, scale=0.4]{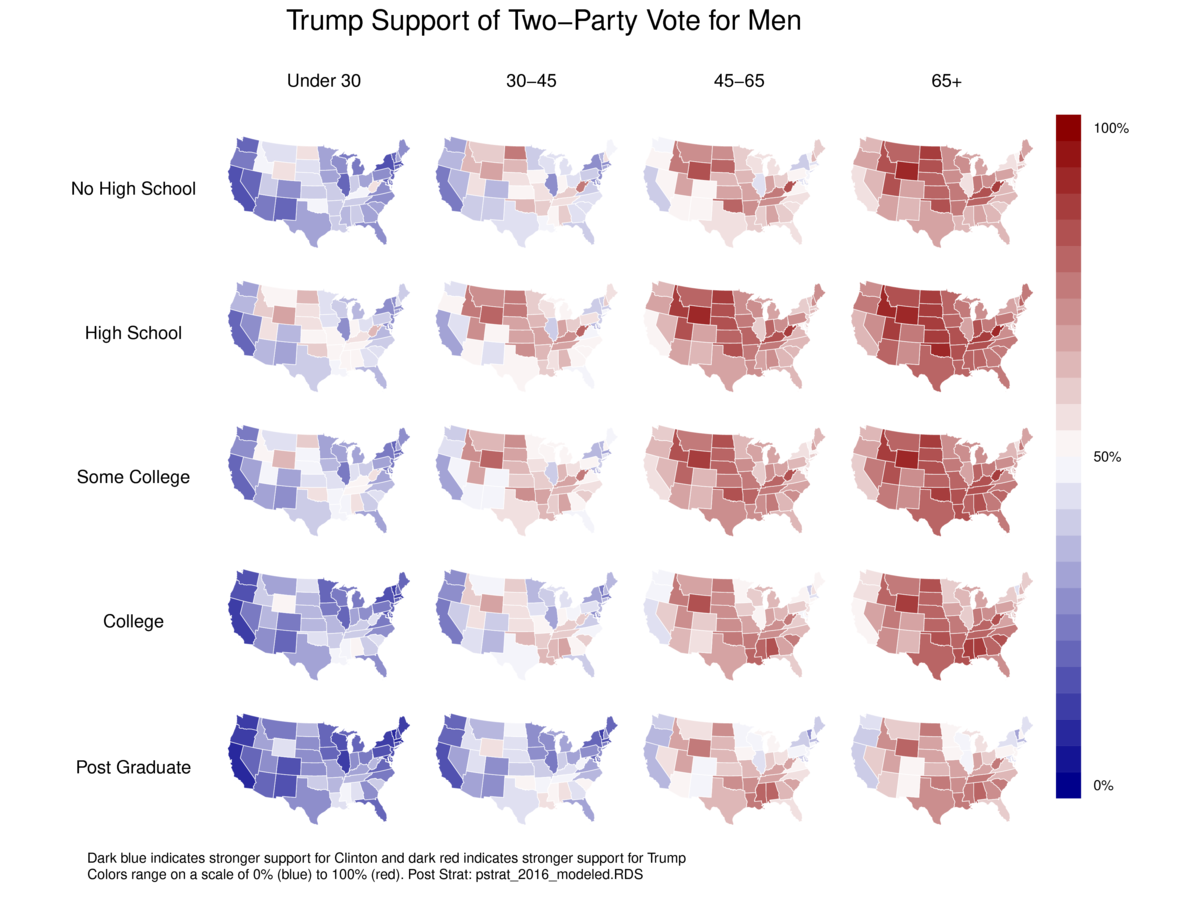}
\footnotesize
\emph{Notes:}  State-level vote intention by education and age for men. Dark red indicates stronger support for Donald Trump and dark blue indicates stronger support for Hillary Clinton. Older men have stronger support for Trump whereas younger men have stronger support for Clinton. Overall, men with a post graduate education have stronger support for Clinton, while men with a high school education have stronger support for Trump. \\
(Using \emph{Model 1}.)
\end{minipage}
\end{figure}

\begin{figure}[H]\caption[]{Trump's Share of the Two-Party Vote by Age and Education for Whites}
\begin{minipage}{1\linewidth}
\includegraphics[trim={0cm 2cm 0cm 2cm}, clip, scale=0.4]{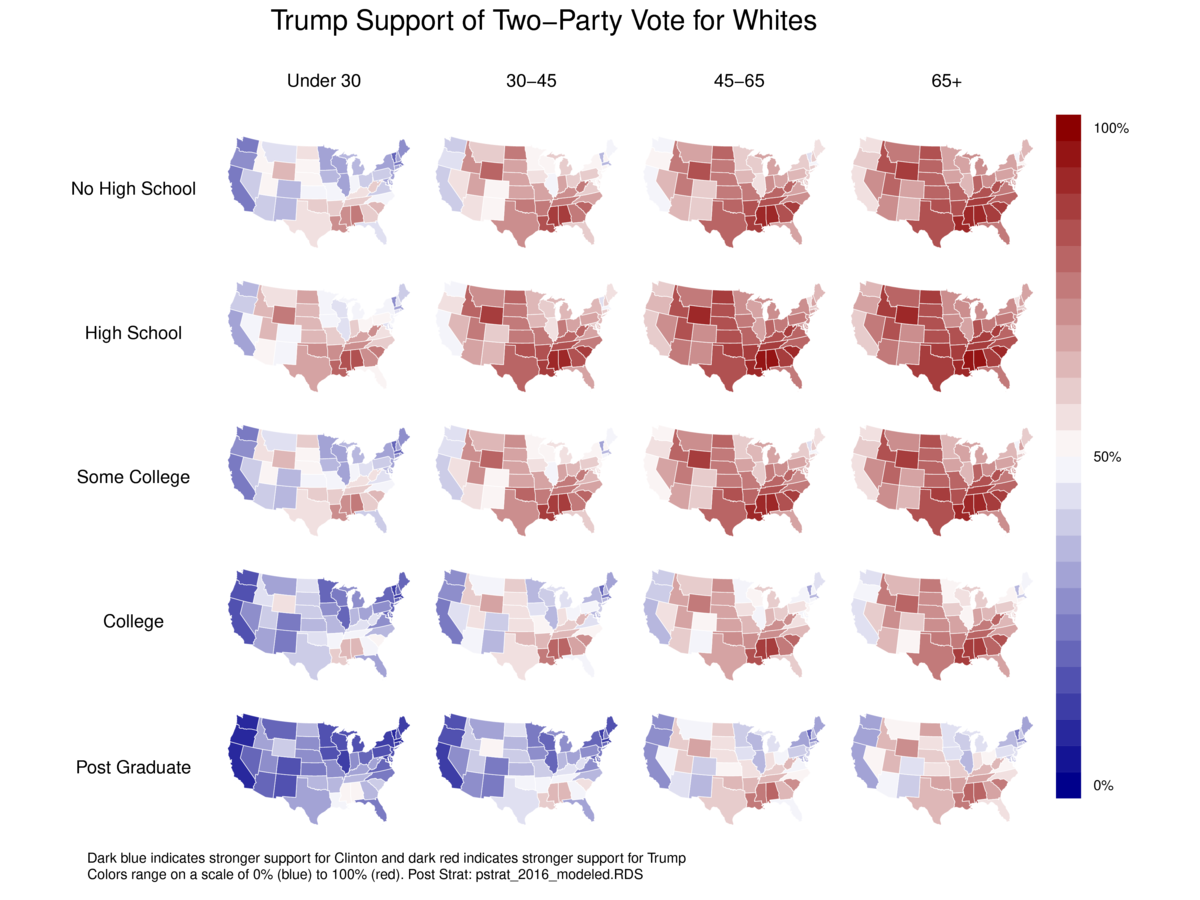}
\footnotesize
\emph{Notes:} State-level vote intention by education and age for Whites. Dark red indicates stronger support for Donald Trump and dark blue indicates stronger support for Hillary Clinton. Older voters with less education had stronger support for Trump, whereas younger voters with more education had stronger support for Clinton. In terms of education, the strongest support for Clinton comes from voters with a post graduate education and the strongest support for Trump comes from voters with a high school education.\\
(Using \emph{Model 1}.)
\end{minipage}
\end{figure}

\begin{figure}[H]\caption[]{Trump's Share of the Two-Party Vote by Age and Education for Blacks}
\begin{minipage}{1\linewidth}
\includegraphics[trim={0cm 2cm 0cm 2cm}, clip, scale=0.4]{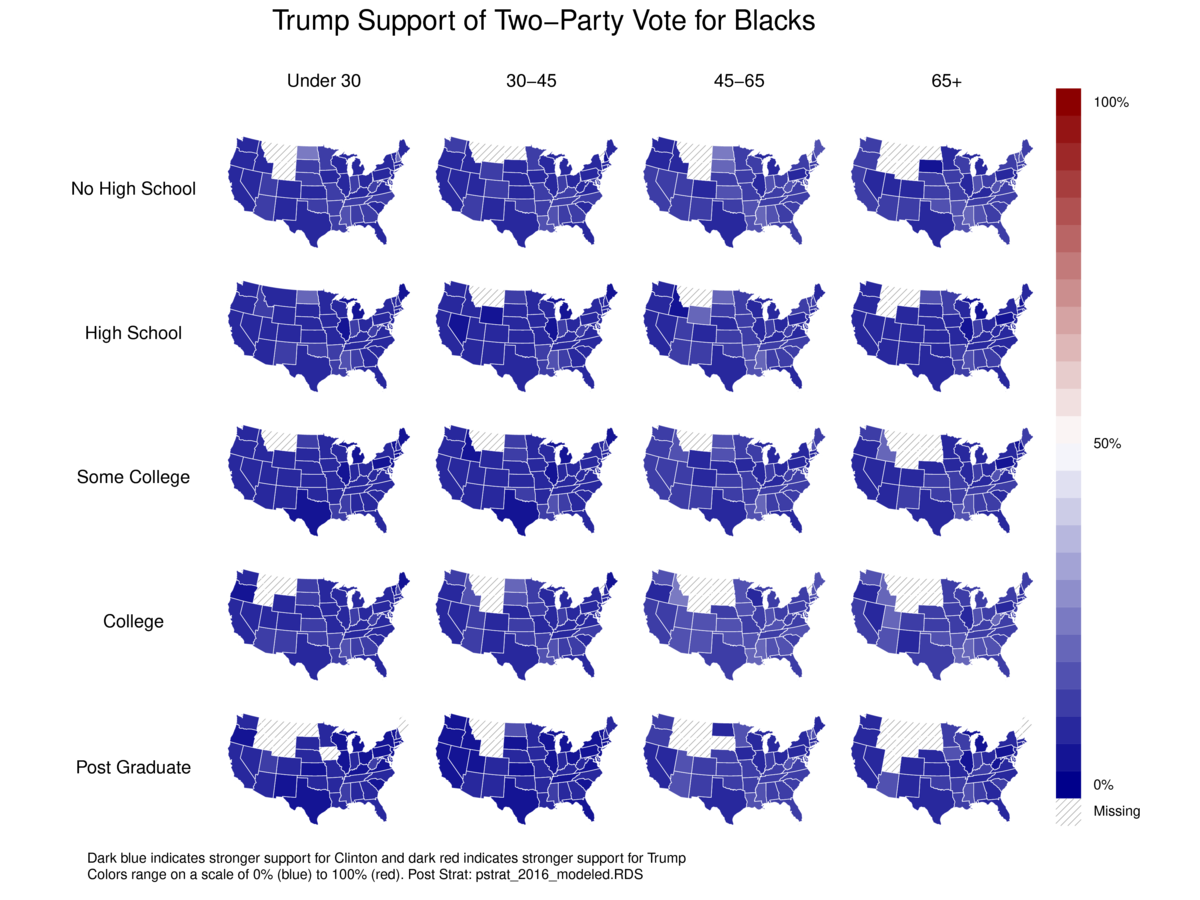}
\footnotesize
\emph{Notes:} State-level vote intention by education and age for Blacks. Dark red indicates stronger support for Donald Trump among women and dark blue indicates stronger support for Hillary Clinton. Missing cells are denoted by diagonal lines. Overall, Blacks supported Clinton. \\
(Using \emph{Model 1}.)
\end{minipage}
\end{figure}

\begin{figure}[H]\caption[]{Trump's Share of the Two-Party Vote by Age and Education for Hispanics}
\begin{minipage}{1\linewidth}
\includegraphics[trim={0cm 2cm 0cm 2cm}, clip, scale=0.4]{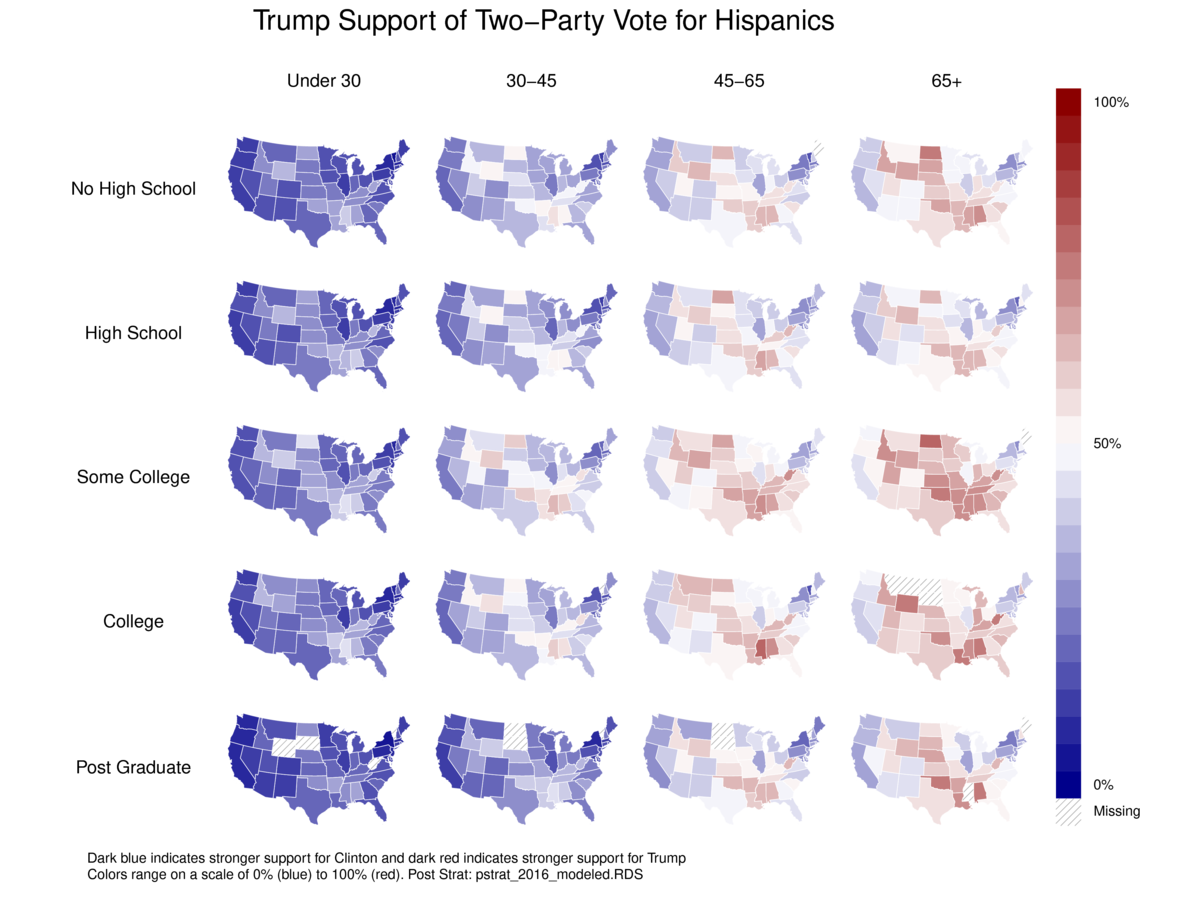}
\footnotesize
\emph{Notes:} State-level vote intention by education and age for Hispanics. Dark red indicates stronger support for Donald Trump and dark blue indicates stronger support for Hillary Clinton. Missing cells are denoted by diagonal lines. A majority of young Hispanics have stronger support for Clinton. Support for Trump increases with age at all education levels. There is not much variation across education levels.  \\
(Using \emph{Model 1}.)
\end{minipage}
\end{figure}

\begin{figure}[H]\caption[]{Trump's Share of the Two-Party Vote by Age and Education for Other Ethnicities}
\begin{minipage}{1\linewidth}
\includegraphics[trim={0cm 2cm 0cm 2cm}, clip, scale=0.4]{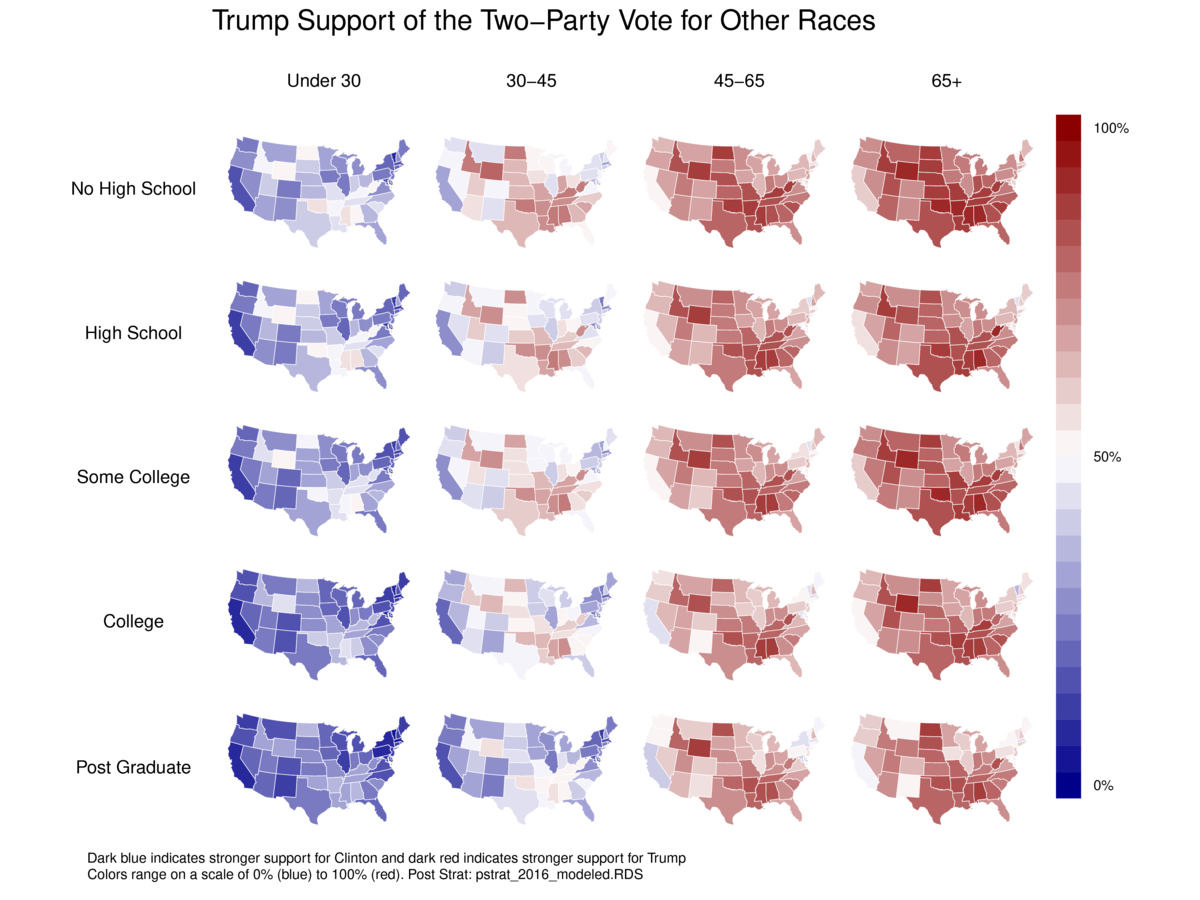}
\footnotesize
\emph{Notes:} State-level vote intention by education and age for ethnicities (not including White, Black, or Hispanic). Dark red indicates stronger support for Donald Trump and dark blue indicates stronger support for Hillary Clinton. Support for Trump increases with age at all education levels. Support for Trump consistently decreases with education (with the exception of the 65+ age bracket). \\
(Using \emph{Model 1}.)
\end{minipage}
\end{figure}

\begin{figure}[H]\caption[]{Trump's Share of the Two-Party Vote by Education and White vs. Non-white}
\begin{center}
\begin{minipage}{0.75\linewidth}
\includegraphics[trim={6cm 5cm 0cm 6.65cm}, clip, scale=0.35]{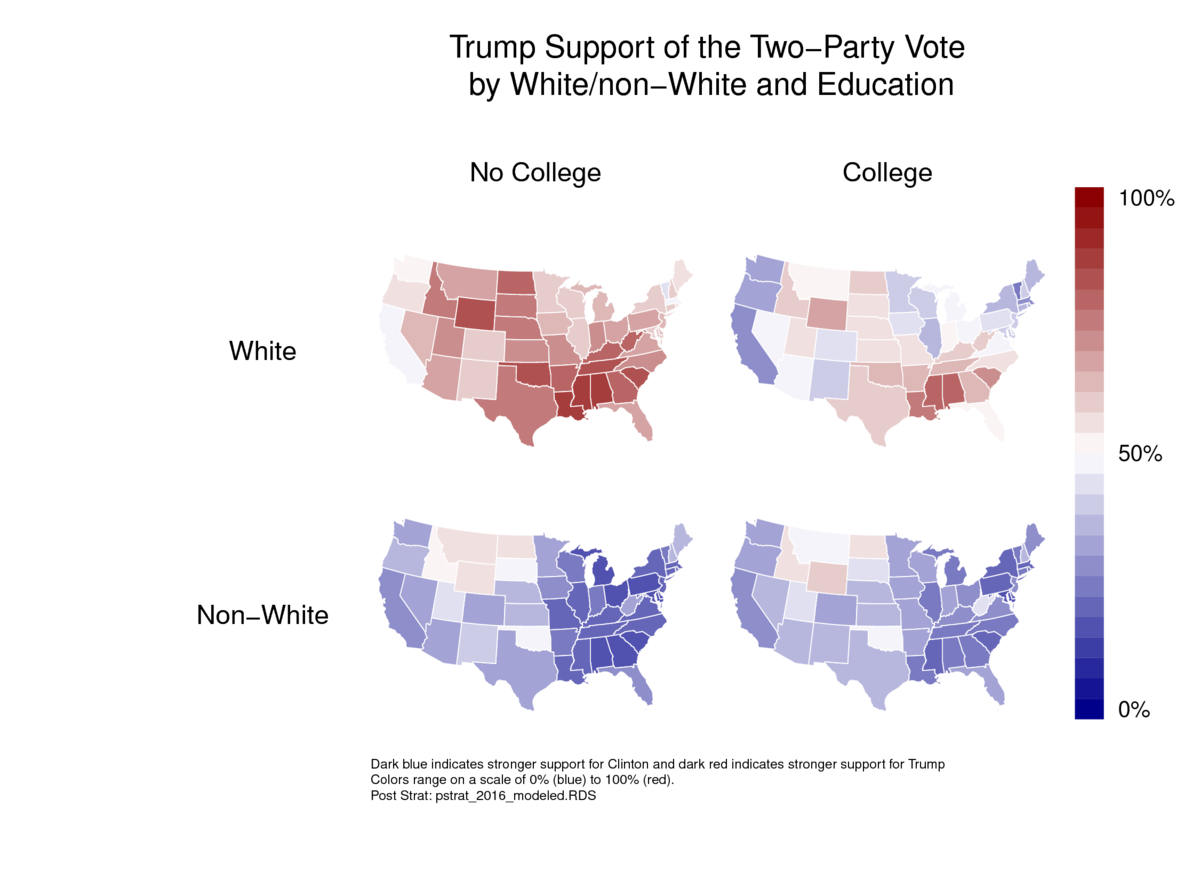}
\footnotesize
\emph{Notes:} State-level vote intention for white and non-white voters by education. No college education includes the categories ``No High School", ``High School", and ``Some College". College education includes the categories ``College" and ``Post Graduate". Dark red indicates stronger support for Donald Trump and dark blue indicates stronger support for Hillary Clinton. White voters have stronger support for Trump compared to non-white voters, with white voters with no college education having the strongest support. There is little variation in vote preference across these categories for North Dakota, Wyoming, and Idaho, which consistently support Trump. There is also little variation in vote preference across education levels among non-white voters.\\
(Using \emph{Model 1}.)
\end{minipage}
\end{center}
\end{figure}
\begin{figure}[H]\caption[]{Romney's Share of the Two-Party Vote by Education and White vs. Non-white - 2012 Election}
\begin{center}
\begin{minipage}{0.7\linewidth}
\includegraphics[trim={6cm 5cm 0cm 6.65cm}, clip, scale=0.35]{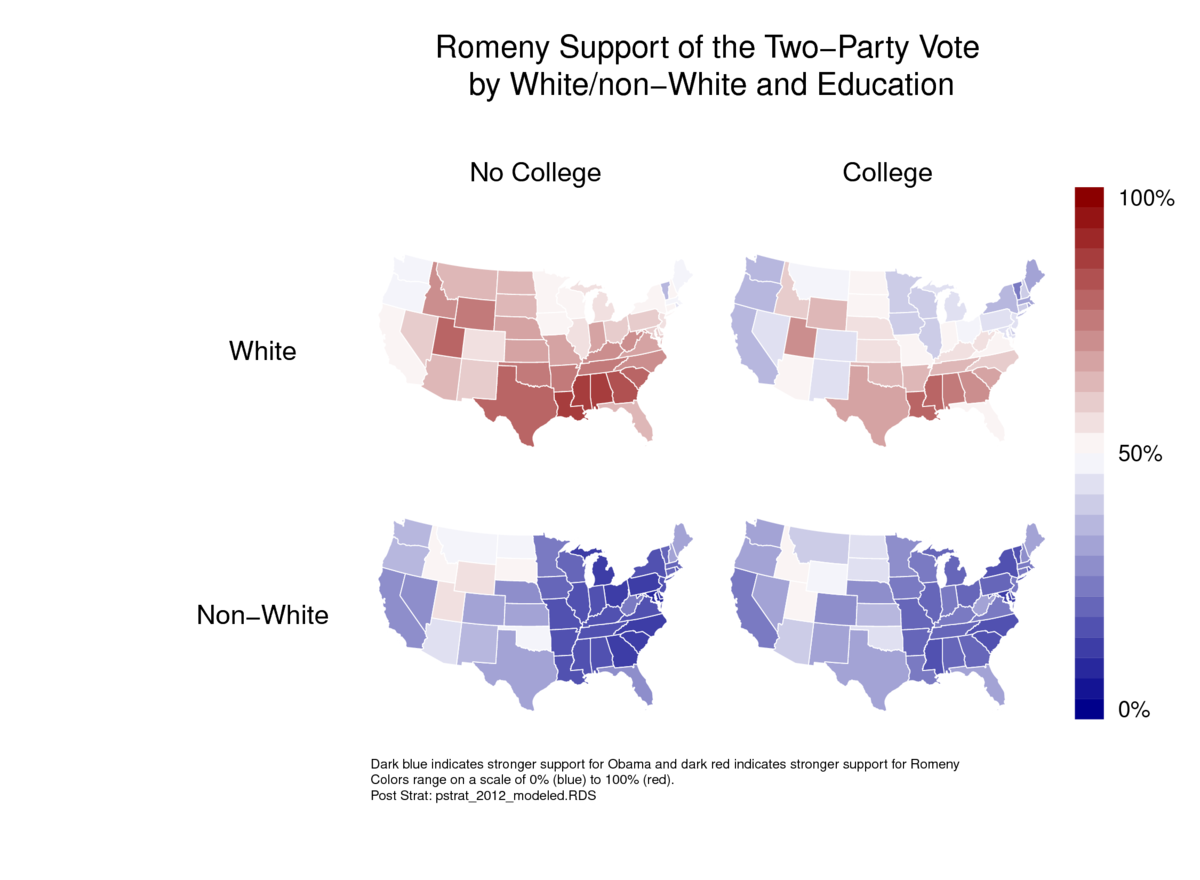}
\footnotesize
\emph{Notes:} State-level vote intention for white and non-white voters by education. No college education includes the categories ``No High School", ``High School", and ``Some College". College education includes the categories ``College" and ``Post Graduate". Dark red indicates stronger support for Mitt Romney and dark blue indicates stronger support for Barack Obama. White voters with no college education had the strongest support for Romney. Regardless of college education, non-White voters had the strongest support for Obama. \\
(Using \emph{Model 1} with 2012 election results/turnout data.)
\end{minipage}
\end{center}
\end{figure}

\begin{figure}[H]\caption[]{Trump's Share of the Two-Party Vote by Education and White vs. Non-white Women}
\begin{center}
\begin{minipage}{1\linewidth}
\includegraphics[trim={2cm 2cm 2cm 2cm}, clip, scale=0.42]{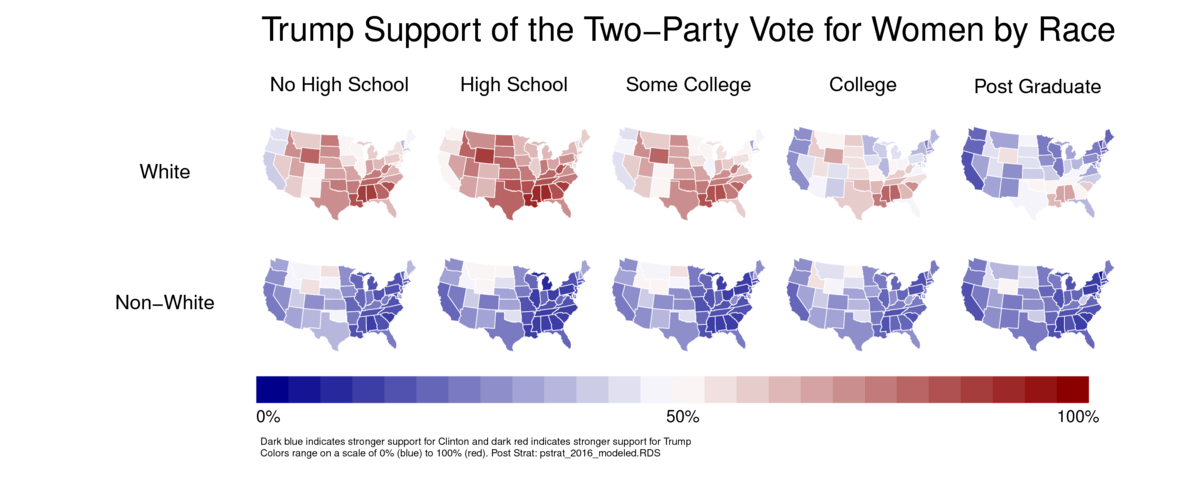}
\footnotesize
\emph{Notes:} State-level vote intention for white and non-white women by education. Dark red indicates stronger support for Donald Trump among women and dark blue indicates stronger support for Hillary Clinton among women. Support for Trump among white women increases from no high school to high school education levels and declines from high school to post graduate education levels. White women with high school education have the strongest support for Trump. Overall, non-white women have stronger support for Clinton, with the exception of some Midwestern states (e.g. North Dakota and Wyoming). \\
(Using \emph{Model 1}.)
\end{minipage}
\end{center}
\end{figure}
\begin{figure}[H]\caption[]{Romney's Share of the Two-Party Vote by Education and White vs. Non-white Women - 2012 Election}
\begin{center}
\begin{minipage}{1\linewidth}
\includegraphics[trim={2cm 2cm 2cm 2cm}, clip, scale=0.42]{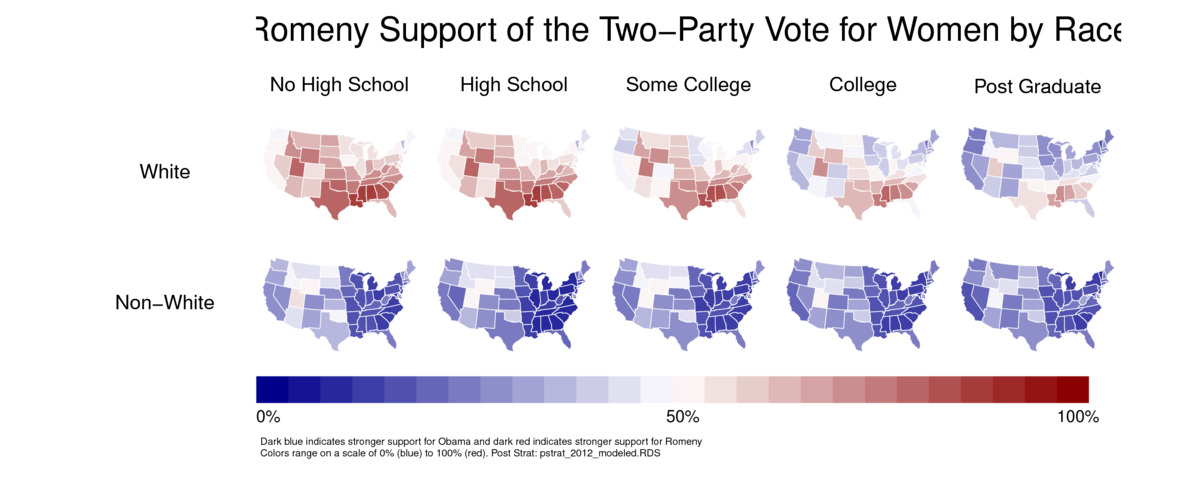}
\footnotesize
\emph{Notes:} State-level vote intention for white and non-white women by education. Dark red indicates stronger support for Mitt Romney among women and dark blue indicates stronger support for Barack Obama among women. Support for Romney among White women decreased with education. Regardless of college education, Obama had strong support among non-White women.\\
(Using \emph{Model 1} with 2012 election results/turnout data.)
\end{minipage}
\end{center}
\end{figure}

\begin{figure}[H]\caption[]{Trump's Share of the Two-Party Vote by Education for Women}
\begin{center}
\begin{minipage}{1\linewidth}
\includegraphics[trim={2cm 2.5cm 2cm 3cm}, clip, scale=0.45]{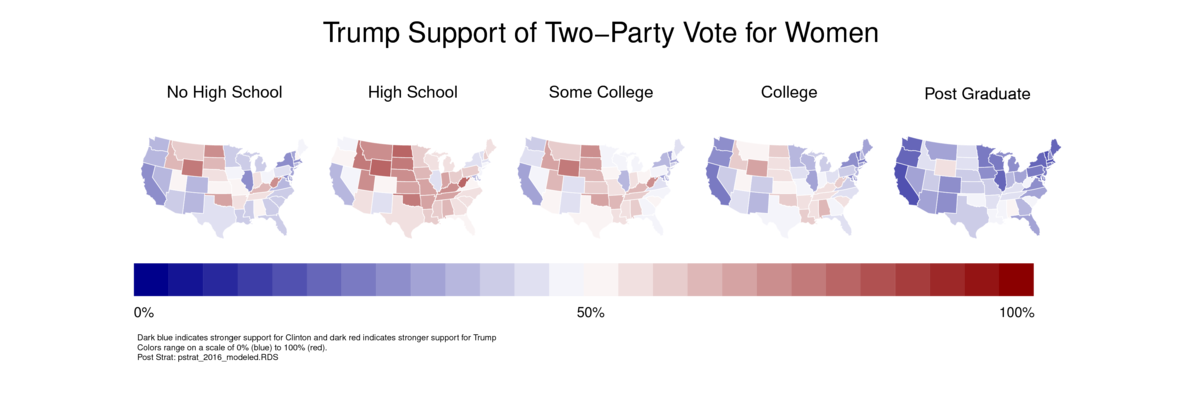}
\footnotesize
\emph{Notes:} State-level vote intention for women by education. Dark red indicates stronger support for Donald Trump among women and dark blue indicates stronger support for Hillary Clinton among women. In most states, women with high school education have stronger support for Trump and women with post graduate education have stronger support for Clinton.  \\
(Using \emph{Model 1}.)
\end{minipage}
\end{center}
\end{figure}
\begin{figure}[H]\caption[]{Romney's Share of the Two-Party Vote by Education for Women - 2012 Election}
\begin{center}
\begin{minipage}{1\linewidth}
\includegraphics[trim={2cm 2.5cm 2cm 3cm}, clip, scale=0.45]{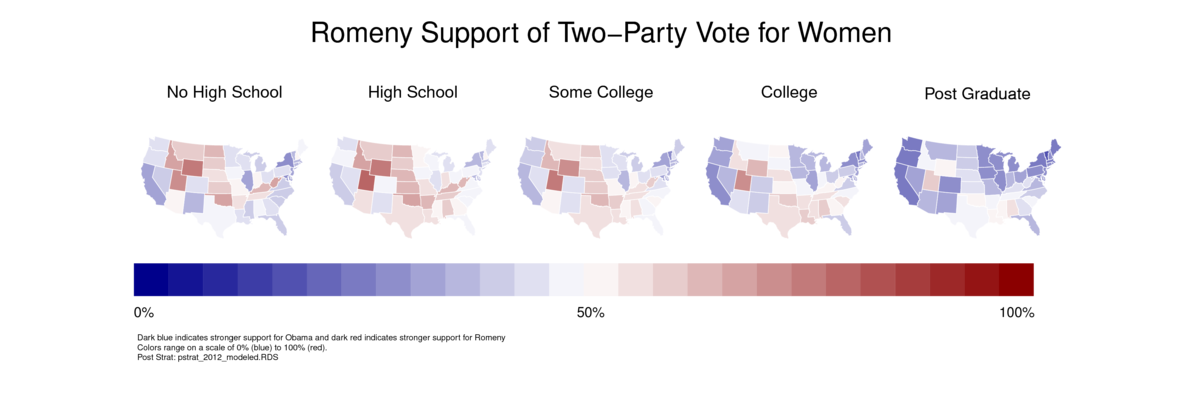}
\footnotesize
\emph{Notes:} State-level vote intention for women by education. Dark red indicates stronger support for Mitt Romney among women and dark blue indicates stronger support for Barack Obama among women. In most states, women with high school education had stronger support for Romney and women with post graduate education had stronger support for Obama.  \\
(Using \emph{Model 1} with 2012 election results/turnout data.)
\end{minipage}
\end{center}
\end{figure}

\subsubsection{Maps of voter turnout}

\begin{figure}[H]\caption[]{Voter Turnout by Age and Education}
\begin{minipage}{1\linewidth}
\includegraphics[trim={0cm 2cm 0cm 2cm}, clip, scale=0.4]{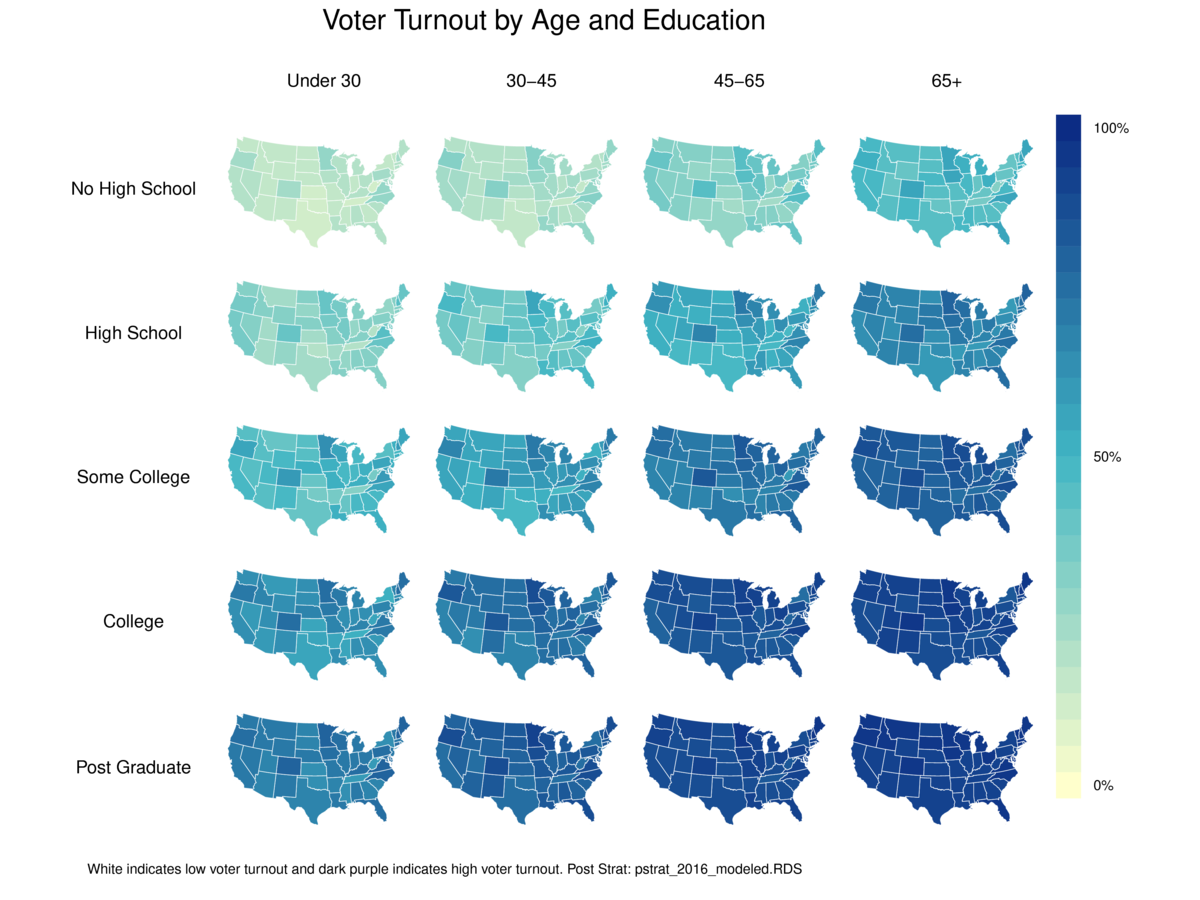}
\footnotesize
\emph{Notes:} State-level voter turnout by education and age. Yellow indicates low voter turnout and dark blue indicates high voter turnout. Younger individuals with less education were less likely to vote this election, whereas older individuals with more education were more likely to vote. \\
(Using \emph{Model 1}.)
\end{minipage}
\end{figure}

\begin{figure}[H]\caption[]{Voter Turnout by Age and Education for Women}
\begin{minipage}{1\linewidth}
\includegraphics[trim={0cm 2cm 0cm 2cm}, clip, scale=0.4]{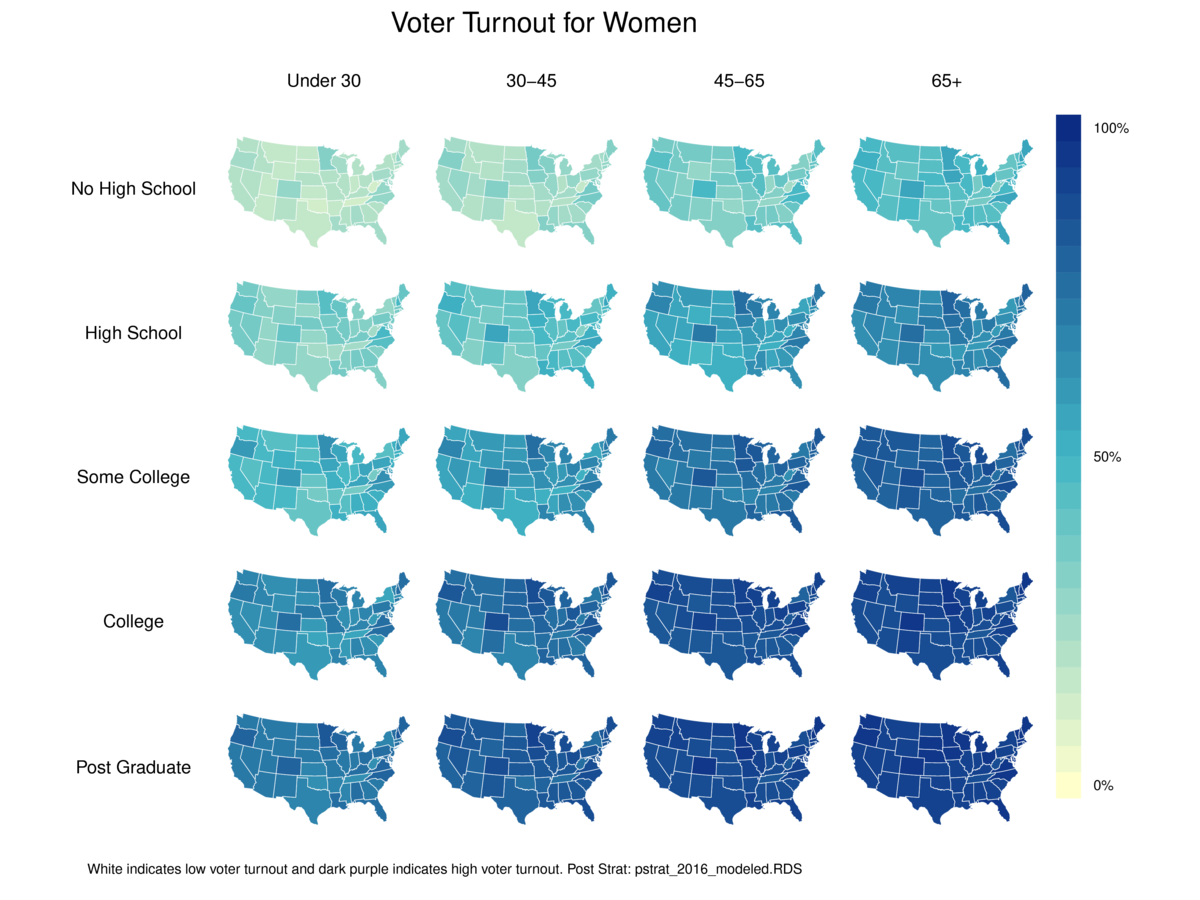}
\footnotesize
\emph{Notes:} State-level voter turnout by education and age for women. Yellow indicates low voter turnout and dark blue indicates high voter turnout. \\
(Using \emph{Model 1}.)
\end{minipage}
\end{figure}

\begin{figure}[H]\caption[]{Voter Turnout by Age and Education for Men}
\begin{minipage}{1\linewidth}
\includegraphics[trim={0cm 2cm 0cm 2cm}, clip, scale=0.4]{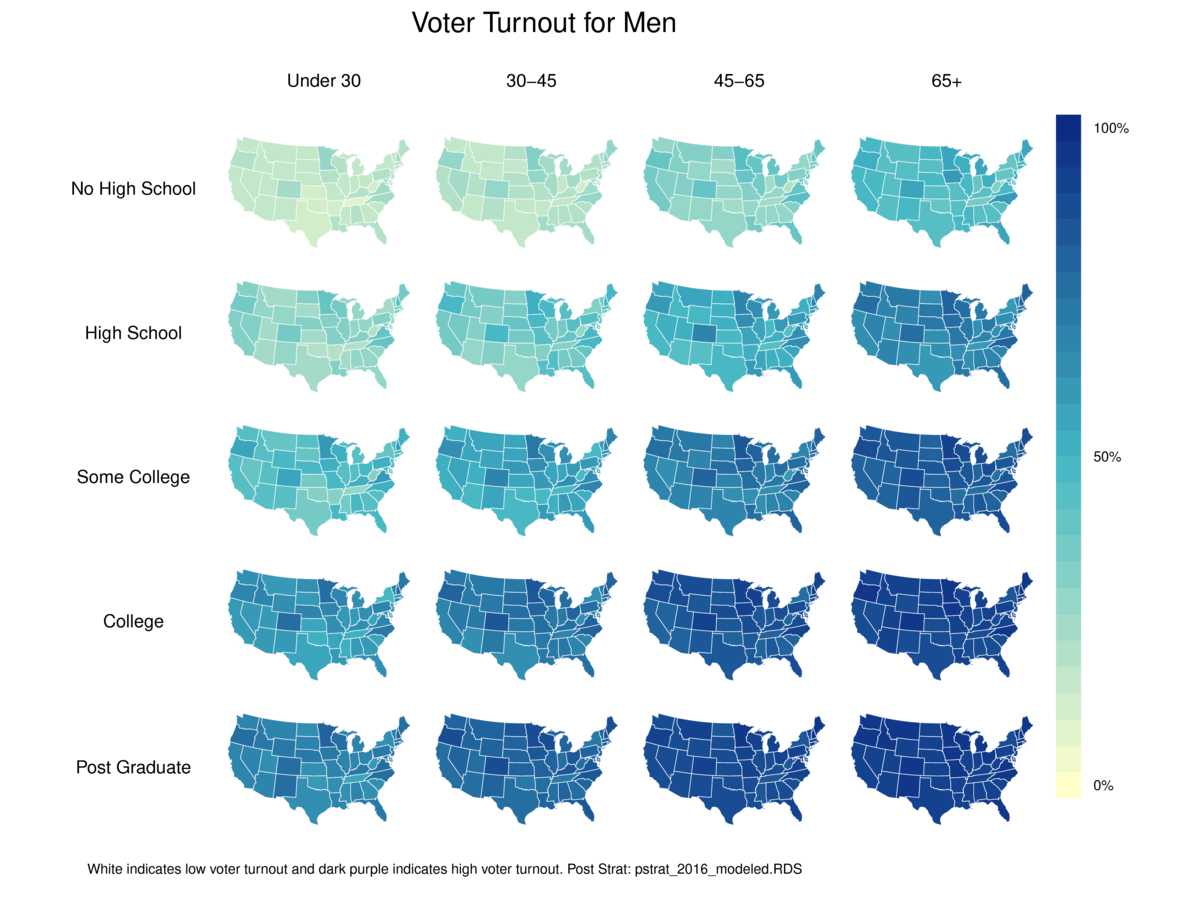}
\footnotesize
\emph{Notes:} State-level voter turnout by education and age for women. Yellow indicates low voter turnout and dark blue indicates high voter turnout. \\
(Using \emph{Model 1}.)
\end{minipage}
\end{figure}

\begin{figure}[H]\caption[]{Voter Turnout Gender Gap (men minus women)}
\begin{minipage}{1\linewidth}
\includegraphics[trim={0cm 2cm 0cm 2cm}, clip, scale=0.4]{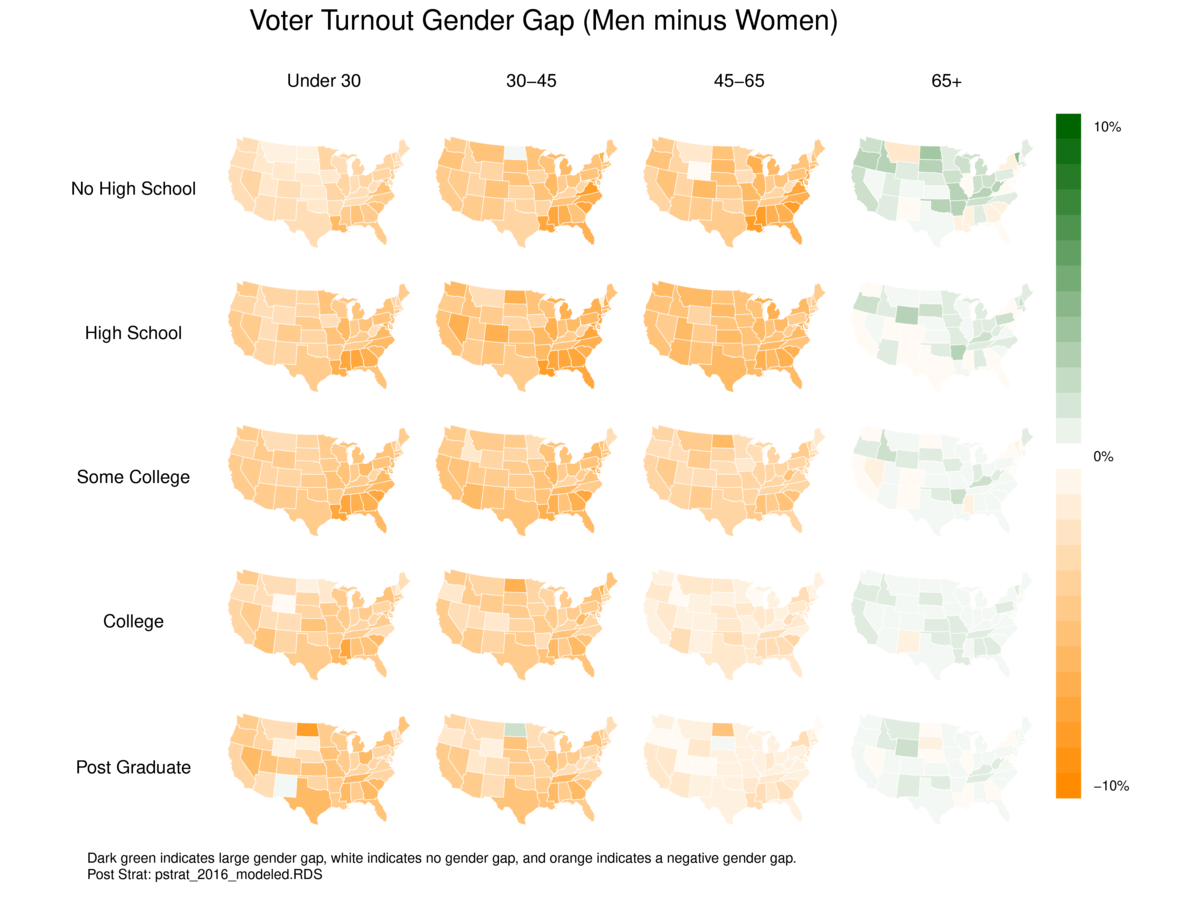}
\footnotesize
\emph{Notes:} State-level voter turnout gender gap evaluated as voter turnout probability for men minus voter turnout probability for women. Dark green/orange indicates a large turnout gender gap. \\
(Using \emph{Model 1}.)
\end{minipage}
\end{figure}


\section{Discussion}

We keep the discussion short as we feel that our main contribution here is to
present these graphs and maps which others can interpret how they see best,
and to share our code so that others can fit these and similar models on their own.

Some of our findings comport with the broader media narrative developed in the
aftermath of the election. We found that white voters with lower educational
attainment supported Trump nearly uniformly. We did not find that income was a
strong predictor of support for Trump, perhaps a continuation of a trend
apparent in 2000 through 2012 election data. We found the gender gap to be
about 10\%, which was a bit lower than predicted by exit polls. The marital
status gap we estimated was about 2$\times$ the figure estimated by exit
polls.

Most surprising to us was the strong age pattern in the gender gap. Older women
were much more likely to support Clinton than older men, while younger women
were mildly more likely to support Clinton compared to men the same age. We are
not sure what accounts for this difference. One area of future research is
using age as a continuous predictor rather than binning ages and using the bins
as categorical predictors.

Our models predict that men in several state by education categories were more
likely to support Clinton than women. We do not believe this to be true but
rather believe it to be a problem with poststratification table sparsity. In
order to reduce the number of poststratification cells, in future analyses we
could poststratify by region rather than state. This would likely not have
impacted our descriptive precision in this analysis due to the apparently
strong regional patterns in voting behavior in this election.

\newpage
\section{Appendix A - Model Code}

We specified our voter turnout model as below:

\begin{verbatim}
cbind(vote, did_not_vote) ~ 1 + female + state_pres_vote +
      (1 | state) + (1 | age) +
      (1 | educ) + (1 + state_pres_vote | eth) +
      (1 | marstat) + (1 | marstat:age) +
      (1 | marstat:state) + (1 | marstat:eth) +
      (1 | marstat:gender) + (1 | marstat:educ) +
      (1 | state:gender) + (1 | age:gender) +
      (1 | educ:gender) + (1 | eth:gender) +
      (1 | state:eth) + (1 | state:age) +
      (1 | state:educ) + (1 | eth:age) +
      (1 | eth:educ) + (1 | age:educ) +
      (1 | state:educ:age) + (1 | educ:age:gender)
\end{verbatim}

We specified our voter preference model as below:

\begin{verbatim}
cbind(clinton, trump) ~ 1 + female + state_pres_vote +
      (1 | state) + (1 | age) +
      (1 | educ) + (1 + state_pres_vote | eth) +
      (1 | marstat) + (1 | marstat:age) +
      (1 | marstat:state) + (1 | marstat:eth) +
      (1 | marstat:gender) + (1 | marstat:educ) +
      (1 | state:gender) + (1 | age:gender) +
      (1 | educ:gender) + (1 | eth:gender) +
      (1 | state:eth) + (1 | state:age) +
      (1 | state:educ) + (1 | eth:age) +
      (1 | eth:educ) + (1 | age:educ) +
      (1 | state:educ:age) + (1 | educ:age:gender)
\end{verbatim}

\newpage
\nocite{*}
\bibliographystyle{unsrt}
\bibliography{mrp_election_bib}

\end{document}